\pdfoutput=1
\documentclass[%
 aip,
 amsmath,amssymb,
reprint,%
]{revtex4-1}

\usepackage{graphicx}% Include figure files
\usepackage{dcolumn}% Align table columns on decimal point
\usepackage{bm}% bold math
\usepackage[utf8]{inputenc}
\usepackage[T1]{fontenc}
\usepackage{mathptmx}
\usepackage{etoolbox}

\makeatletter
\def\@email#1#2{%
 \endgroup
 \patchcmd{\titleblock@produce}
  {\frontmatter@RRAPformat}
  {\frontmatter@RRAPformat{\produce@RRAP{*#1\href{mailto:#2}{#2}}}\frontmatter@RRAPformat}
  {}{}
}%
\makeatother
\begin{document}

% \preprint{AIP/123-QED}

% \title[Sample title]{Sample Title:\\with Forced Linebreak}
% Force line breaks with \\

\title{Split Electrons in Partition Density Functional Theory }

\author{Kui Zhang}
\affiliation{ 
Department of Physics and Astronomy, Purdue university, West Lafayette, IN 47907 USA
}%

\author{Adam Wasserman}%
 \email{awasser@purdue.edu}
\affiliation{ 
Department of Physics and Astronomy, Purdue university, West Lafayette, IN 47907 USA
}%
\affiliation{
Department of Chemistry, Purdue university, West Lafayette, IN 47907 USA
}%

\date{\today}% It is always \today, today,

\begin{abstract}
	Partition Density Functional Theory (P-DFT) is a density embedding method that partitions a molecule into fragments by minimizing the sum of fragment energies subject to a local density constraint and a global electron-number constraint. To perform this minimization, we study a two-stage procedure in which the sum of fragment energies is lowered when electrons flow from fragments of lower electronegativity to fragments of higher electronegativity. The global minimum is reached when all electronegativities are equal. The non-integral fragment populations are dealt with in two different ways: (1) An {\em ensemble} approach (ENS) that involves averaging over calculations with different numbers of electrons (always integers); and (2) A simpler approach that involves fractionally occupying orbitals (FOO). We compare and contrast these two approaches and examine their performance in some of the simplest systems where one can transparently apply both, including simple models of heteronuclear diatomic molecules and actual diatomic molecules with 2 and 4 electrons. We find that, although both ENS and FOO methods lead to the same total energy and density, the ENS fragment densities are less distorted than those of FOO when compared to their isolated counterparts, and they tend to retain integer numbers of electrons. We establish the conditions under which the ENS populations can become fractional and observe that, even in those cases, the total charge transferred is always lower in ENS than in FOO. Similarly, the FOO fragment dipole moments provide an upper bound to the ENS dipoles. We explain why, and discuss implications.

\end{abstract}

\maketitle

%Kui: Please fix all captions so that they accurately describe the figures.

%%%%%%%%%%%
%INDRODUCTION%
%%%%%%%%%%%
\section{Introduction}
\label{sec:intro}

Kohn-Sham Density Functional Theory (KS-DFT) 
\cite{hohenberg_inhomogeneous_1964, kohn_self-consistent_1965,parr_density-functional_1989}, continues to be one of the most powerful and widely used methods to calculate the 
electronic properties of matter. Approximate exchange-correlation (XC) functionals used within KS-DFT suffer from various errors that limit its applicability \cite{cohen_challenges_2012}. Among these errors, delocalization and static-correlation errors have been widely studied \cite{cohen_insights_2008, cohen_challenges_2012} but, in spite of several recent approaches to correct them 
%Kui: add here a few appropriate references
%\cite{flosic, orbitalets, kraisler_elimination_2015, recent-machine-learning,others}
\cite{pederson_communication_2014,kraisler_elimination_2015,nafziger_fragment-based_2015,li_localized_2018,perdew_artificial_2021,kirkpatrick_pushing_2021}, there is still no general, robust method that works reliably in practice. 

%The origin of the ubiquitous delocalization error can be traced back to the incorrect treatment of fractional numbers of electrons. 
When dissociating a molecule into its constituent atoms, most approximate XC functionals will minimize the energy by placing fractional numbers of electrons on the separated atoms. The differences between these fractional numbers and the correct integers, referred to here as the ``fractional-charge error" (FCE), are sometimes taken as a measure of the delocalization error (DE) 
%Kui: Please cite appropriate Yang paper
\cite{mori-sanchez_localization_2008}.  But the DE, whose origin is the incorrect delocalization of electron densities, is not only present at dissociation. It is also present at any set of internuclear separations, for which the concept of `atomic electron number' is no longer valid (i.e. there is no unique way to assign electrons to any particular nucleus in the molecule). Although the FCE is only well defined at dissociation, one can  show (see for example Figure \ref{fig:ABp_disso}) that the error in the energy associated to the FCE settles in long before reaching dissociation when the `atoms' are still at interacting distances and their electronic densities cannot be understood separately. Theories that provide a definite prescription for calculating fragment populations at finite internuclear separations are thus useful in this context \cite{cohen_hardness_2006,tang_fragment_2012,fabiano_frozen_2014,schulz_description_2019}. 
%They provide the means to link the DE and the FCE at {\em any} set of internuclear separations.

Partition-DFT (P-DFT) \cite{cohen_foundations_2007, elliott_partition_2010, nafziger_density-based_2014} is a formally exact density embedding method that leads to such a definite prescription, and can thus be used to link the DE and the FCE at {\em any} set of internuclear separations. There are two sensible ways for treating fractional electron numbers in P-DFT\cite{nafziger_accurate_2017,jiang_constructing_2018}: (1) An {\em ensemble} approach (ENS) that involves averaging over calculations with different numbers of electrons (always integers); and (2) A simpler approach that involves fractionally occupying KS orbitals (FOO). In this work, we compare and contrast these two approaches and examine their performance in some of the simplest systems where one can transparently apply both.

All of the P-DFT calculations of fractional charges done to date \cite{cohen_charge_2009, elliott_density_2009, tang_fragment_2012, oueis_exact_2018, nafziger_density-based_2014, nafziger_accurate_2017} have been 
carried out on model systems of non-interacting \cite{cohen_charge_2009, elliott_density_2009, tang_fragment_2012} or one-dimensional interacting \cite{oueis_exact_2018} electrons. Whenever 3D P-DFT calculations have involved fractional charges, the molecules studied have been centro-symmetric (homonuclear diatomics), not involving ground-state charge transfer between the constituent atoms \cite{nafziger_density-based_2014, nafziger_accurate_2017}.
%An extension to calculations for real heteronuclear diatomic molecules would be desirable. 
%But it raises new challenges because spatial symmetry cannot be used and
%the fragment electron populations need to be optimized.
We present here the first calculations on 3D heteronuclear diatomic molecules and ions. Our goal is to examine how the two methods for splitting electrons between fragments (ENS and FOO) differ in practice. In Sec.II, we review the theory highlighting the role of electronegativity equalization and show how this important condition plays differently in FOO and ENS. In Sec.III we present P-DFT calculations on 3D heteronuclear diatomic molecules with 1, 2, and 4 electrons, and show that, even though both ENS and FOO methods lead to the same total molecular densities, they lead to different descriptions of the charge transferred between fragments.   

%Furthermore, in P-DFT, FOO and ENS result in 
%different partitionings for the same total density of the system.
%The nonadditive noninteracting kinetic energies and potentials
%for homonuclear diatomic molecules 
%produced by FOO and ENS were compared 
%\cite{nafziger_accurate_2017}.
%For heteronuclear diatomic molecules, FOO and ENS 
%may give different optimal electron populations.

%In the following, we first formulate the P-DFT with constrained electron populations of fragments. 
%In this way, by changing fragment electron populations, we can find the optimal ones
%and charge transfer can be defined. 
%Then we give P-DFT with FOO and ENS, separately, which are used to deal with non-integer 
%electron numbers.
%Furthermore, we discuss the relationship bewteen HOMO energy and asymptotic behaviour of density.
%Next, we do numerical calculation for a one-electron and a two-electron heteronuclear diatomic model molecules to compare FOO and ENS.
%Finally, we also do P-DFT calculation for real helium hydride ion
%and lithium hydride.

%%%%%%%%%%%%%%%%%%
%  COMPUTATIONAL  DETAILS  %
%%%%%%%%%%%%%%%%%%
%%
\section{Partition-DFT and two alternative methods for determining fractional electron numbers}
% \label{sec:CD}

%%
To highlight the role of fractional populations in P-DFT, we begin with a description of P-DFT in which the $\{N_{\alpha}\}$ ($\alpha$ is used to label fragments) are presumed fixed and known in advance. We will then describe how to optimize the set $\{N_{\alpha}\}$, but - as a first step - we consider these numbers as given, for example, by a calculation of formal charges from any of the many methods available for that purpose 
\cite{mulliken_electronic_1955,hirshfeld_bonded-atom_1977,reed_natural_1985,bader_atoms_1990,rousseau_atomic_2001}.
%Kui: Please add 3-4 references here.
%\cite{Mulliken-charges, Bader, Hirshfeld, Voronoi...}.

%The illustrations of the next section will be done in systems composed of two fragments, but we begin by describing Partition-DFT \cite{elliott_partition_2010} in a more general case that can be applied to molecular systems made of an arbitrary number of fragments, $N_f$.
%\subsection{P-DFT with Constrained Fragment Electron Populations}
%\label{ssec:vp}
%In partition theory (PT), we first divide 
For a system of $N_{M}$ electrons subject to an external potential $v\left(\mathbf{r}\right)$ (due to all the nuclei), this potential  can be divided as
%$The total `external' potential (due to all the nuclei)  %$v\left(\mathbf{r}\right)$ is divided as:
%of a system into $N_{f}$ fragments:
\begin{equation}
v\left(\mathbf{r}\right)=\underset{\alpha=1}{\overset{N_{f}}{\sum}}v_{\alpha}\left(\mathbf{r}\right),
\end{equation}
where $v_{\alpha}\left(\mathbf{r}\right)$ is the external potential of the $\alpha^{\rm th}$ fragment and $N_f$ is the number of fragments.
The task of P-DFT is to minimize the sum of fragment energies
\begin{equation}
\label{e:Ef}
E_{f}\left[\left\{ n_{\alpha}\right\} \right]\equiv\underset{\alpha=1}{\overset{N_{f}}{\sum}}E_{\alpha}\left[n_{\alpha}\right]
\end{equation}
subject to the set of constraints specified below (Eqs.(\ref{density_conserved}-\ref{N_alpha_conserved})).  In Eq.(\ref{e:Ef}), $E_{\alpha}\left[n_{\alpha}\right]$ and $n_{\alpha}\left(\mathbf{r}\right)$ are the ground-state energy and density of the $\alpha^{\rm th}$ fragment. Given a set $\{N_\alpha\}$ satisfying $\underset{\alpha=1}{\overset{N_{f}}{\sum}}N_{\alpha}=N_{M}$, where the total number of electrons $N_M$ is an integer, the total ground-state density $n_{M}\left(\mathbf{r}\right)$ is optimally partitioned 
when $E_f$ is minimized with respect to variations of the $\left\{ n_{\alpha}\right\}$ subject to the $N_f+1$ independent constraints: 
\begin{equation}
\label{density_conserved}
n_{M}\left(\mathbf{r}\right)=\underset{\alpha=1}{\overset{N_{f}}{\sum}}n_{\alpha}\left(\mathbf{r}\right),
\end{equation}
\begin{equation}
\label{N_alpha_conserved}
N_{\alpha}=\intop n_{\alpha}\left(\mathbf{r}\right)d\mathbf{r},\; \forall\alpha\in\left\{ 1,\cdots,N_{f}\right\}.
\end{equation}
%with $N_{\alpha}$ being the electron population of the $\alpha$th fragment;
%$N_{M}$ 
%Though $N_{M}$ is an integer, $N_{\alpha}$ can be non-integral.

Because the $N_\alpha$ can be non-integer,
$E_{\alpha}\left[n_{\alpha}\right]$ needs to be defined for non-integer electron numbers. Two alternative methods for accomplishing this are described in Sections II-A and II-B. In either case, the constrained search for an optimum $\{n_\alpha\}$ for fixed $\{N_\alpha\}$ can be formally converted into the unconstrained minimization of a grand-potential $G\left[\left\{ n_{\alpha}\right\} ,v_{p}\left(\mathbf{r}\right),\left\{ \chi_{\alpha}\right\} \right]$, where the {\em partition potential} $v_p(\mathbf{r})$ and fragment electronegativities $\left\{ \chi_{\alpha}\right\}$ are the Lagrange multipliers associated to constraints (\ref{density_conserved}) and (\ref{N_alpha_conserved}), respectively:  
\begin{eqnarray}
\nonumber
G && \left[\left\{ n_{\alpha}\right\},v_{p}\left(\mathbf{r}\right),\left\{ \chi_{\alpha}\right\} \right]=
E_{f}\left[\left\{ n_{\alpha}\right\} \right]+\int v_{p}\left(\mathbf{r}\right)\times\\
&&\left(\underset{\alpha}{\sum}n_{\alpha}\left(\mathbf{r}\right)-n_{M}\left(\mathbf{r}\right)\right)d\mathbf{r}+\underset{\alpha}{\sum}\chi_{\alpha}\left(\int n_{\alpha}\left(\mathbf{r}\right)d\mathbf{r}-N_{\alpha}\right).
\label{grand-potential}
\end{eqnarray}

The Euler-Lagrange equation for the $\alpha^{\rm th}$ fragment can be obtained 
by minimizing 
$G\left[\left\{ n_{\alpha}\right\} ,v_{p}\left(\mathbf{r}\right),\left\{ \chi_{\alpha}\right\} \right]$
with respect to the corresponding fragment density $n_{\alpha}\left(\mathbf{r}\right)$, i.e., $\frac{\delta G}{\delta n_{\alpha}\left(\mathbf{r}\right)}=0$, 
yielding
\begin{equation}
\label{Euler_equ}
\chi_{\alpha} = 
-\left( \frac{\delta E_{\alpha}\left[n_{\alpha}\right]}{\delta n_{\alpha}\left(\mathbf{r}\right)}+v_{p}\left(\mathbf{r}\right) \right),
\end{equation}
where $\frac{\delta E_{\alpha}\left[n_{\alpha}\right]}{\delta n_{\alpha}\left(\mathbf{r}\right)}$
needs to be well defined for non-integer electron numbers (see Secs.II-A and II-B).

The $v_{p}\left(\mathbf{r}\right)$ for fixed $\{N_\alpha\}$ is unique
\cite{cohen_hardness_2006, cohen_foundations_2007,huang_quantum_2011}, but $E_f$ can generally be lowered further by transferring a fraction of an electron from a fragment of lower electronegativity to one of higher electronegativity. The global minimum of $E_f$ is then found when all fragment electronegativities are equal \cite{cohen_hardness_2006, cohen_foundations_2007},
i.e. $\chi_{\alpha}=\chi^{*},\forall\alpha $, where $\chi^{*}$ is a common optimal electronegativity.
Then the constraints (\ref{N_alpha_conserved}) unify into:
\begin{equation}
\label{N_conserved}
N_{M}=\underset{\alpha=1}{\overset{N_{f}}{\sum}}\intop n_{\alpha}\left(\mathbf{r}\right)d^{3}\mathbf{r} ,
\end{equation}
and the grand potential 
$G\left[\left\{ n_{\alpha}\right\} ,v_{p}\left(\mathbf{r}\right),\left\{ \chi_{\alpha}\right\} \right]$ 
becomes
\begin{eqnarray}
\nonumber
G  && \left[\left\{ n_{\alpha}\right\} ,v_{p}\left(\mathbf{r}\right),\chi^*\right]=
E_{f}\left[\left\{ n_{\alpha}\right\} \right]+\int v_{p}\left(\mathbf{r}\right)\times\\
&& \left(\underset{\alpha}{\sum}n_{\alpha}\left(\mathbf{r}\right)-n_{M}\left(\mathbf{r}\right)\right)d^{3}\mathbf{r}+\chi^*\left(\underset{\alpha}{\sum}\int n_{\alpha}\left(\mathbf{r}\right)d^{3}\mathbf{r}-N_{M}\right)
\end{eqnarray}
with $\chi^*=-\left(\frac{\delta E_{f}\left[\left\{ n_{\alpha}\right\} \right]}{\delta n_{M}\left(\mathbf{r}\right)}+v_{p}\left(\mathbf{r}\right)\right)$.
 
Thus,
by varying the $\left\{ N_{\alpha}\right\} $, we can determine the 
optimal fragment electron populations
for which all fragment electronegativities are equal and the grand-potential is minimized to:
\begin{equation}
\label{minimize_G}
E_{f}^{*}=\underset{\left\{ \chi_{\alpha}\right\} }{\min}\underset{\left\{ n_{\alpha}\right\} }{\min}G\left[\left\{ n_{\alpha}\right\} ,v_{p}\left(\mathbf{r}\right),\left\{ \chi_{\alpha}\right\} \right]
\end{equation}
%According to the electronegativity equalization principle \cite{sanderson_interpretation_1951},
%we can define each fragment as one "atomic" component of the molecule.
%It is obvious that 
%%\begin{equation}
%%\label{minimize_G}
%%G^*\left[\left\{ n_{\alpha}\right\} ,v_{p}\left(\mathbf{r}\right),\chi_{M}\right]=
%%\underset{\left\{ \chi_{\alpha}\right\} }{\min}G\left[\left\{ n_{\alpha}\right\} ,v_{p}\left(\mathbf{r}\right),\left\{ \chi_{\alpha}\right\} \right].
%%\end{equation}
%Considering that for 
%$G\left[\left\{ n_{\alpha}\right\} ,v_{p}\left(\mathbf{r}\right),\chi_{M}\right]$
%all fragment electronegativities are equal,
A general, simple algorithm for Eq.(\ref{minimize_G}) can be formulated:
Start with a guess for $\left\{ N_{\alpha}\right\} $
and do self-consistent calculations to obtain all fragment electronegativities 
$\left\{ \chi_{\alpha}\right\} $.
Then update the $\left\{ N_{\alpha}\right\} $ 
by using  
$N_{\alpha}^{\left(k+1\right)}=N_{\alpha}^{\left(k\right)}
+\Gamma\left(\chi_{\alpha}^{\left(k\right)}-\bar{\chi}^{\left(k\right)}\right)$
\cite{elliott_partition_2010}, 
where $\Gamma$ is an appropriate positive constant and $\bar{\chi}^{(k)}$ is the average of 
all fragment electronegativities for the $k^{\rm th}$ iteration. The process is iterated until $\chi_{\alpha}^{\left(k\right)}-\bar{\chi}^{\left(k\right)}$ falls below an acceptable threshold. The resulting $\{N_\alpha\}$ will be referred to as {\em optimal} and be denoted as $\{N_\alpha^*\}$ from now on. We now turn our attention to two alternative methods for treating non-integer electron numbers.
\subsection{Fractional Orbital populations (FOO)}
\label{ssec:numerical}

The density of the $\alpha^{\rm th}$ fragment can be constructed as
\begin{equation}
n_{\alpha}^{\rm FOO}\left(\mathbf{r}\right)=\underset{i}{\sum}f_{\alpha,i}\left|\phi_{\alpha,i}\left(\mathbf{r}\right)\right|^{2},
\label{n_FOO}
\end{equation}
where $\underset{i}{\sum}f_{\alpha,i}=N_{\alpha}$, 
and $f_{\alpha,i}$ is the occupation number for 
$\phi_{\alpha,i}\left(\mathbf{r}\right)$ (the $i^{\rm th}$ KS orbital of the $\alpha^{\rm th}$ fragment).
Since we only consider the ground state, $f_{\alpha,i}=1$ for all occupied orbitals
except for the highest occupied molecular orbital (HOMO), in which case $0<f_{\alpha}^{\rm HOMO}\leq 1$.
The non-interacting kinetic energy is calculated by
\cite{janak_proof_1978}
\begin{equation}
T_{s}^{\mathrm{FOO}}\left[n_{\alpha}\right]=-\frac{1}{2}\underset{i}{\sum}f_{\alpha,i}\int\phi_{\alpha,i}^{*}\left(\mathbf{r}\right)\nabla^{2}\phi_{\alpha,i}\left(\mathbf{r}\right)d\mathbf{r}.
\end{equation}
The energy of the $\alpha^{\rm th}$ fragment $E_{\alpha}\left[n_{\alpha}\right]$ is then defined in the same way as in KS-DFT (we drop the ``FOO" superscript from the densities for notational simplicity),
\begin{equation}
E_{\alpha}\left[n_{\alpha}\right]=T_{s}^{\mathrm{FOO}}\left[n_{\alpha}\right]+
%E_{\mathrm{H}}\left[n_{\alpha}\right]+
E_{\mathrm{HXC}}\left[n_{\alpha}\right]+\int v_{\alpha}\left(\mathbf{r}\right)n_{\alpha}\left(\mathbf{r}\right)d\mathbf{r},
\end{equation}
where $E_{\mathrm{H}}\left[n_{\alpha}\right]$ and $E_{\mathrm{XC}}\left[n_{\alpha}\right]$
are the Hartree and exchange-correlation energies of the $\alpha^{\rm th}$ fragment.
Then the Euler-Lagrange equation  (\ref{Euler_equ}) becomes
\begin{equation}
\chi_{\alpha}^{\rm FOO}=
-\left( \frac{T_{s}^{\mathrm{FOO}}\left[n_{\alpha}\right]}{\delta n_{\alpha}\left(\mathbf{r}\right)}+v_{\mathrm{HXC}}\left[n_{\alpha}\right]\left(\mathbf{r}\right)
%+v_{\mathrm{XC}}\left[n_{\alpha}\right]\left(\mathbf{r}\right)
+v_{\alpha}\left(\mathbf{r}\right)+v_{p}\left(\mathbf{r}\right) \right).
\end{equation}

% \subsubsection{fragment KS equation}
% for Ef, adding Lagrange multipliers according to constraints, then do functional derivation respect to orbitals 
In the same way as in KS-DFT \cite{parr_density-functional_1989}, 
one can derive the KS equations for the $\alpha^{\rm th}$ fragment:
\begin{equation}
\label{KS-FOO}
\left\{ -\frac{1}{2}\nabla^{2}+v_{\alpha}^{\mathrm{eff}}\left[n_{\alpha}\right]\left(\mathbf{r}\right)+v_{p}\left(\mathbf{r}\right)\right\} \phi_{\alpha,i}\left(\mathbf{r}\right)=\epsilon_{\alpha,i}\phi_{\alpha,i}\left(\mathbf{r}\right),
\end{equation}
where the fragment effective potential is
\begin{equation}
\label{v_eff}
v_{\alpha}^{\mathrm{eff}}\left[n_{\alpha}\right]\left(\mathbf{r}\right)=v_{\alpha}\left(\mathbf{r}\right)+v_{\mathrm{HXC}}\left[n_{\alpha}\right]\left(\mathbf{r}\right).
%+v_{\mathrm{XC}}\left[n_{\alpha}\right]\left(\mathbf{r}\right).
\end{equation}

The fragment KS equations for the $\alpha^{\rm th}$ fragment
can be regarded as those for $N_{\alpha}$ electrons subject to the external potential 
$v_{\alpha}\left(\mathbf{r}\right)+v_{p}\left(\mathbf{r}\right)$.
Therefore, we define the total energy for the $\alpha^{\rm th}$ fragment as 
\begin{equation}
\label{total_frag_energy}
\widetilde{E}_{\alpha}\left[n_{\alpha}\right]\equiv E_{\alpha}\left[n_{\alpha}\right]+\int v_{p}\left(\mathbf{r}\right)n_{\alpha}\left(\mathbf{r}\right)d\mathbf{r}.
\end{equation}
Considering that $N_{\alpha}$ is continuous and the energy is differentiable, 
we use the definition of electronegativity given by 
Iczkowski and Margrave \cite{iczkowski_electronegativity_1961}, 
i.e., $\chi_{\alpha}=-\frac{\partial\widetilde{E}_{\alpha}}{\partial N_{\alpha}}$.
It is straightforward to prove that the energy functional $\widetilde{E}_{\alpha}\left[n_{\alpha}\right]$
is differentiable \cite{englisch_exact_1984, englisch_exact_1984-1}
and 
$\chi_{\alpha}^{\rm FOO}=-\frac{\delta\widetilde{E}_{\alpha}\left[n_{\alpha}\right]}{\delta n_{\alpha}\left(\mathbf{r}\right)}=-\frac{\partial\widetilde{E}_{\alpha}}{\partial N_{\alpha}}$
\cite{parr_density-functional_1989, ripka_quantum_1986}.
Furthermore, according to Janak's theorem \cite{janak_proof_1978}, we know that 
$\frac{\partial\widetilde{E}_{\alpha}}{\partial N_{\alpha}}=\epsilon_{\alpha}^{\mathrm{HOMO}}$.
Thus, we obtain $\ensuremath{\chi_{\alpha}^{\rm FOO}}=-\epsilon_{\alpha}^{\mathrm{HOMO}}$.

\subsection{Ensembles (ENS)}

Alternatively, when $N_{\alpha}$ lies between the integers $p_\alpha$ ($=\left\lfloor N_{\alpha}\right\rfloor $) and $p_{\alpha+1}$, we can write the $\alpha$-density as:
\begin{equation}
\label{n_alpha}
n_{\alpha}^{\rm ENS}\left(\mathbf{r}\right)=\left(1-\omega_{\alpha}\right)n_{p_{\alpha}}\left(\mathbf{r}\right)+\omega_{\alpha}n_{p_{\alpha}+1}\left(\mathbf{r}\right),
\end{equation}
emulating known results for the exact extension of DFT to non-integer electron numbers \cite{perdew_density-functional_1982} (true for the {\em exact} XC-functional). A related approach \cite{kraisler_elimination_2015} has been shown to be useful in eliminating the asymptotic fractional dissociation problem of the Local Density Approximation (LDA). In Eq.(\ref{n_alpha}), the two ensemble component densities $n_{p_{\alpha}}\left(\mathbf{r}\right)$ and $n_{p_{\alpha+1}}\left(\mathbf{r}\right)$ integrate to $p_\alpha$ and $p_{\alpha+1}$ electrons, respectively, and
%For the $\alpha$th fragment, its electron population can always be written as
%\begin{equation}
%\label{N_alpha}
%N_{\alpha}=p_{\alpha}+\omega_{\alpha}=\left(1-\omega_{\alpha}\right)p_{\alpha}+\omega_{\alpha}\left(p_{\alpha}+1\right),
%\end{equation}
%where $p_{\alpha}$ is a non-negative integer and 
$0<\omega_{\alpha}<1$ so that $N_\alpha = p_\alpha + \omega_\alpha$.

%According to ENS \cite{perdew_density-functional_1982},
%the ensemble describing the fragment is 
%a mixture of a $p_{\alpha}$-electron pure state $\Psi_{p_{\alpha}}$ 
%with probability $1-\omega_{\alpha}$
%and a ($p_{\alpha}+1$)-electron pure state $\Psi_{p_{\alpha}+1}$
%with probability $\omega_{\alpha}$.
%Therefore, the density matrix of the $\alpha$th fragment is
%$\hat{\rho}_{\alpha}=\left(1-\omega_{\alpha}\right)\left|\Psi_{p_{\alpha}}\right\rangle \left\langle \Psi_{p_{\alpha}}\right|+\omega_{\alpha}\left|\Psi_{p_{\alpha}+1}\right\rangle \left\langle \Psi_{p_{\alpha}+1}\right|$.
%In this way, the density is given by

The corresponding fragment energy is
\begin{equation}
E_{\alpha}\left[n_{\alpha}\right]=\left(1-\omega_{\alpha}\right)E_{\alpha}\left[n_{p_{\alpha}}\right]+\omega_{\alpha}E_{\alpha}\left[n_{p_{\alpha}+1}\right],
\end{equation}
where $E_{\alpha}\left[n_{p_{\alpha}}\right]$ and 
$E_{\alpha}\left[n_{p_{\alpha}+1}\right]$
are the ground-state energies for $p_{\alpha}$ and $p_{\alpha}+1$ electrons 
subject to an external potential 
$v_{\alpha}\left(\mathbf{r}\right)+v_{p}\left(\mathbf{r}\right)$.
% $E_{\alpha}\left[n_{\alpha}\right]$ cannot be expressed directly 
% in the form of KS-DFT using $n_{\alpha}$. 

% So we do not have an explicit expression for fragment chemical potentials. 
Since the allowed density variations must now keep $p_\alpha$ electrons in $n_{p_\alpha}(\mathbf{r})$ and $p_{\alpha+1}$ electrons in $n_{p_{\alpha+1}}(\mathbf{r})$, the constraints (\ref{N_alpha_conserved}) turn into $2N_f$ independent constraints, and the Lagrange multipliers $\{\chi_\alpha\}$ can be replaced by a pair-set of multipliers $\{\lambda_{p_\alpha},\lambda_{p_\alpha+1}\}$.  The grand-potential in Eq.(\ref{grand-potential}), now $G^{\mathrm{ENS}}\left[\left\{ n_{\alpha}\right\} ,v_{p}\left(\mathbf{r}\right), \left\{ \lambda_{p_{\alpha}},\lambda_{p_{\alpha}+1}\right\} \right]$, is minimized with respect to variations of the $n_{p_\alpha}(\mathbf{r})$ and $n_{p_{\alpha+1}}(\mathbf{r})$, yielding
\begin{equation}
\lambda_{p_{\alpha}}=
-\left( \frac{\delta E_{\alpha}\left[n_{p_{\alpha}}\right]}{\delta n_{p_{\alpha}}\left(\mathbf{r}\right)}+v_{p}\left(\mathbf{r}\right)\right),
\end{equation}
% In the same way, we can obtain the Euler-Lagrange equation for the $(p_{\alpha}+1)$-electron element of the ensemble for the $\alpha$th fragment:
% \begin{equation}
% \frac{\delta E_{\alpha}\left[n_{p_{\alpha}+1}\right]}{\delta n_{p_{\alpha}+1}\left(\mathbf{r}\right)}+v_{p}\left(\mathbf{r}\right)=\mu_{p_{\alpha}+1}.
% \end{equation}
and a companion equation for $\lambda_{p_\alpha+1}$. Expressing $E_{\alpha}\left[n_{p_{\alpha}}\right]$ in terms of KS quantities, Eq.(\ref{KS-FOO}) is now replaed by a pair of analogous KS equations in which all of the $\alpha$-subindices in Eq.(\ref{KS-FOO}) are replaced by either $p_\alpha$ or $p_\alpha +1$.  
%\begin{equation}
%\lambda_{p_{\alpha}}=
%-\left( \frac{\delta T_{s}\left[\left\{ n_{p_{\alpha}}\right\} \right]}{\delta n_{p_{\alpha}}\left(\mathbf{r}\right)}+v_{\mathrm{H}}\left[n_{p_{\alpha}}\right]\left(\mathbf{r}\right)+v_{\mathrm{XC}}\left[n_{p_{\alpha}}\right]\left(\mathbf{r}\right)+v_{\alpha}\left(\mathbf{r}\right)+v_{p}\left(\mathbf{r}\right) \right).
%\end{equation}
% \subsubsection{KS equations for ensemble components}
%In the same way as KS-DFT, we can derive the KS equations for 
%the $p_{\alpha}$-electron element of the ensemble for the $\alpha$th fragment:
%%\begin{equation}
%%\left\{ -\frac{1}{2}\nabla^{2}+v_{p_{\alpha}}^{\mathrm{eff}}\left[n_{p_{\alpha}}\right]\left(\mathbf{r}\right)+v_{p}\left(\mathbf{r}\right)\right\} \phi_{p_{\alpha},i}\left(\mathbf{r}\right)=\epsilon_{p_{\alpha},i}\phi_{p_{\alpha},i}\left(\mathbf{r}\right)
%%\end{equation}
%%together with the corresponding equation for $p_{\alpha+1}$.
The fragment effective potential $v_{p_{\alpha}}^{\mathrm{eff}}\left[n_{p_{\alpha}}\right]\left(\mathbf{r}\right)$ has the same decomposition as in Eq.(\ref{v_eff}), but the Hartree and XC-potentials are now evaluated at the appropriate integer-number densities. These in turn are constructed simply by summing over the $p_\alpha$ (or $p_{\alpha+1}$) occupied KS orbitals. 
%where $n_{p_{\alpha}}=\underset{i}{\sum}\left|\phi_{p_{\alpha},i}\left(\mathbf{r}\right)\right|^{2}$, 
%and the fragment effective potential is
%\begin{equation}
%v_{p_{\alpha}}^{\mathrm{eff}}\left[n_{p_{\alpha}}\right]\left(\mathbf{r}\right)=v_{\alpha}\left(\mathbf{r}\right)+v_{\mathrm{H}}\left[n_{p_{\alpha}}\right]\left(\mathbf{r}\right)+v_{\mathrm{XC}}\left[n_{p_{\alpha}}\right]\left(\mathbf{r}\right).
%\end{equation}
%The fragment KS equation can be regarded as $p_{\alpha}$ electrons subject to the external potential $v_{\alpha}\left(\mathbf{r}\right)+v_{p}\left(\mathbf{r}\right)$.

%In the same way, we can obtain the Euler-Lagrange equation and KS equation
%for the $(p_{\alpha}+1)$-electron element of the ensemble for the $\alpha$th fragment.
%Considering that the $p_{\alpha}$-electron and $(p_{\alpha}+1)$-electron element 
%of the ensemble for the $\alpha$th fragment are equivalent, 
%we only need to replace the index $p_{\alpha}$ by $p_{\alpha}+1$ 
%in the above equations.

For atoms and molecules, 
$\epsilon_{p_{\alpha}}^{\mathrm{HOMO}}<\epsilon_{p_{\alpha}+1}^{\mathrm{HOMO}}$ and
$\lambda_{p_{\alpha}}>\lambda_{p_{\alpha}+1}$,
since $p_{\alpha}$ and $p_{\alpha}+1$ electrons are subjected to 
the same external potential $v_{\alpha}\left(\mathbf{r}\right)+v_{p}\left(\mathbf{r}\right)$. 
Defining the total energy for the $\alpha^{\rm th}$ fragment in the same way as
Eq. (\ref{total_frag_energy}), we now find for the electronegativity \cite{perdew_density-functional_1982}:
%It is also straight to prove that for non-integer electron numbers $N_{\alpha}$,
%the energy functionals $\widetilde{E_{\alpha}}\left[n_{\alpha}\right]$
%is differentiable
%\cite{englisch_exact_1984, englisch_exact_1984-1}
%and 
%$\chi_{\alpha}=-\frac{\delta\widetilde{E_{\alpha}}\left[n_{\alpha}\right]}{\delta %n_{\alpha}\left(\mathbf{r}\right)}=-\frac{\partial\widetilde{E_{\alpha}}}{\partial N_{\alpha}}$
%\cite{parr_density-functional_1989, ripka_quantum_1986}.
%Therefore, by using the same definition of electronegativity as in FOO, 
\begin{equation}
\chi_{\alpha}^{\rm ENS}=\widetilde{E}_{\alpha}\left[n_{p_{\alpha}}\right]-\widetilde{E}_{\alpha}\left[n_{p_{\alpha}+1}\right]~~~~~~(0<\omega_\alpha<1).
\label{chi-ENS}
\end{equation}
If $N_\alpha$ is an integer, $\chi_{\alpha}^{\rm ENS}$ is strictly undefined (but one can consider right- and left- limits for it, as will be discussed in Sec.III-B). 

\subsection{HOMO Energy and Electronegativity: FOO {\em vs.} ENS}
%The asymptotic form of the density is determined by HOMO enegy, but not by chemical potential.

The one-electron nature of the KS equations leads to densities for finite systems that decay asymptotically as 
\cite{levy_exact_1984, parr_density-functional_1989}
\begin{equation}
\label{asymptotic_decay}
n\left(\mathbf{r}\right)\underset{r\rightarrow\infty}{\longrightarrow}e^{-2\kappa r},
\end{equation}
where $\kappa=\sqrt{-2\left(\epsilon^{\mathrm{HOMO}}-v_{\mathrm{XC}}\left(\infty\right)\right)}$.
If $v_{\mathrm{XC}}\left(\mathbf{r}\right)\underset{r\rightarrow\infty}{\longrightarrow}0$, 
%that approximate exchange-correlation functionals such as LDA and GGA satisfy, 
then $\kappa=\sqrt{-2\epsilon^{\mathrm{HOMO}}}$.
LDA and GGA densities satisfy this condition.
%, so their exponential decay rate is %Therefore, if we use KS-DFT with LDA or GGA to do calculation for 
%an atom or molecule, 
%the asymptotic decay of the density should be 
%\begin{equation}
%n_{M}\left(\mathbf{r}\right)\underset{r\rightarrow\infty}{\longrightarrow}e^{-2\kappa_{M}r},
%\end{equation}
%where 
%$\kappa_{M}=\sqrt{-2\epsilon_{M}^{\mathrm{HOMO}}}$.

%%In P-DFT, according to the fragment KS equations, 
%%the expression of the asymptotic decay of the density is the same as in
%%Eq.(\ref{asymptotic_decay}) but with %%$\kappa=\sqrt{-2\left(\epsilon^{\mathrm{HOMO}}-v_{\mathrm{XC}}\left(\infty\right)-v_{p}\left(\infty\right)\right)}$.
%%Because $v_{p}\left(\mathbf{r}\right)\underset{r\rightarrow\infty}{\longrightarrow}0$ whenever 
%%$v_{\mathrm{XC}}\left(\mathbf{r}\right)\underset{r\rightarrow\infty}{\longrightarrow}0$, 
%%we find that the P-DFT density also falls asymptotically with the same decay constant $\kappa_M$. Furthermore, 

Due to the density constraint of Eq.(\ref{density_conserved}),  
we see that if the fragment densities have different exponential decay rates, at least one of them (fragment $\alpha^*$)
must  match the decay of the total density in the asymptotic region.  Using FOO, this implies 
%
%For FOO, assume the density of the $\alpha^{*}$th fragment 
%decays in the same way as the total density in asymptotic region, 
%we have 
$\epsilon_{\alpha^{*}}^{\mathrm{HOMO}}=\epsilon_{M}^{\mathrm{HOMO}}=const$.
By using $\chi_{\alpha}^{\rm FOO}=-\epsilon_{\alpha}^{\mathrm{HOMO}}$ (see last line of Sec.II-A), 
we find that $\chi_{\alpha^{*}}^{\rm FOO}=-\epsilon_{M}^{\mathrm{HOMO}}$.
When $E_{f}$ reaches its global minimum, all fragment electronegativities are equal,
so $\chi_{\alpha}=-\epsilon_{M}^{\mathrm{HOMO}}$ for all $\alpha$.
Therefore, the best way to approach the global minimum of $E_{f}$ is 
to always assign less electrons to fragments whose electronegativities are $\chi_{M}$
and more electrons to fragments whose electronegativities are larger than $\chi_{M}$.
In this way, the iteration formula in the algorithm for Eq. (\ref{minimize_G})
can be modified for the $(k+1)^{\rm th}$ iteration as follows: 
For fragments whose $\chi_{\alpha}^{\left(k\right)}>\chi_{M}$, set
$N_{\alpha}^{\left(k+1\right)}=N_{\alpha}^{\left(k\right)}+\Gamma\left(\chi_{\alpha}^{\left(k\right)}-\chi_{M}\right)$;
for fragments whose $\chi_{\alpha}^{\left(k\right)}=\chi_{M}$, set
$N_{\alpha}^{\left(k+1\right)}=N_{\alpha}^{\left(k\right)}$
or
$N_{\alpha}^{\left(k+1\right)}=N_{M}-\underset{\beta\neq\alpha}{\sum}N_{\beta}^{\left(k\right)}$.

% (1. Janak's theorem; 2. at least one fragment whose HOMO energy = HOMO energy of molecule. 
% based on these two points, we can derive that at least one fragment whose chemical potential 
% is equal to the chemical potential of the molecule. therefore, we have the above algorithm.)
% However, fore ENS, we do not have such a algorithm since there is no fragment whose 
% chemical potential is a constant (is equal to the chemical potential of the molecule.)

In contrast, when using the ENS method of Sec.II-B, we find
%$n_{M}=\underset{\alpha=1}{\overset{N_{f}}{\sum}}\left[\left(1-\omega_{\alpha}\right)n_{p_{\alpha}}\left(\mathbf{r}\right)+\omega_{\alpha}n_{p_{\alpha}+1}\left(\mathbf{r}\right)\right]$,
%and 
%$\epsilon_{p_{\alpha}}^{\mathrm{HOMO}}<\epsilon_{p_{\alpha}+1}^{\mathrm{HOMO}}$.
%Thus, 
$\epsilon_{p_{\alpha^{*}+1}}^{\mathrm{HOMO}}=\epsilon_{M}^{\mathrm{HOMO}}=const$, and there is no direct relation between the fragment electronegativity $\chi_\alpha^{\rm ENS}$ of Eq.(\ref{chi-ENS}) and the molecular HOMO energy. 
%Considering that $\epsilon_{p_{\alpha^{*}+1}}^{\mathrm{HOMO}}$ dominates 
%the asymptotic behaviour of density of the $\alpha^{*}$th fragment, 
%we can denote $\epsilon_{\alpha^{*}}^{\mathrm{HOMO}}\equiv\epsilon_{p_{\alpha^{*}+1}}^{\mathrm{HOMO}}$.
%Furthermore, there is no direct relation between fragment electronegativities and HOMO energies.
%Especially, there is no condition to require that  
%$\mu_{\alpha^{*}}$ must be equal to $\epsilon_{p_{\alpha^{*}+1}}^{\mathrm{HOMO}}$,
%which is different from the result of Cohen and Wasserman 
%\cite{cohen_foundations_2007}.

Finally, we note a key difference between the partition potentials of ENS and FOO: Assume the population of the $\alpha^{\rm th}$ fragment is non-integral and $\epsilon_{\alpha,\mathrm{FOO}}^{\mathrm{HOMO}}=\epsilon_{M}^{\mathrm{HOMO}}$, then we know $\varepsilon_{\alpha,\mathrm{ENS}}^{\mathrm{HOMO}}\leq\varepsilon_{M}^{\mathrm{HOMO}}$.
Since there are $N_{\alpha}$ electrons subject to $v_{\alpha}\left(\mathbf{r}\right)+v_{p}^{\mathrm{FOO}}\left(\mathbf{r}\right)$ in FOO
while there are $\left\lfloor N_{\alpha}\right\rfloor +1$ electrons subject to $v_{\alpha}\left(\mathbf{r}\right)+v_{p}^{\mathrm{ENS}}\left(\mathbf{r}\right)$ in ENS, $v_{p}^{\mathrm{ENS}}\left(\mathbf{r}\right)$ needs to be deeper than
$v_{p}^{\mathrm{FOO}}\left(\mathbf{r}\right)$ to account for the fact that
$\left\lfloor N_{\alpha}\right\rfloor +1>N_{\alpha}$,
and thus
$\int v_{p}^{\mathrm{ENS}}\left(\mathbf{r}\right)n_{M}\left(\mathbf{r}\right)d\mathbf{r}<\int v_{p}^{\mathrm{FOO}}\left(\mathbf{r}\right)n_{M}\left(\mathbf{r}\right)d\mathbf{r}$.
As a consequence,
$E_{p}^{\mathrm{ENS}}$ is more negative than $E_{p}^{\mathrm{FOO}}$ and, since FOO and ENS must yield the same total energy, $E_{f}^{\mathrm{ENS}}$ will be less negative
than $E_{f}^{\mathrm{FOO}}$.
All of these findings are illustrated below.
%%Finally, we note a key difference between the partition potentials of ENS and FOO: When the population of the $\alpha^{\rm th}$ fragment is non-integral and  $\epsilon_{\alpha,\mathrm{FOO}}^{\mathrm{HOMO}}=\epsilon_{\alpha,\mathrm{ENS}}^{\mathrm{HOMO}}=\epsilon_{M}^{\mathrm{HOMO}}$, then,
%%since there are $N_{\alpha}$ electrons subject to $v_{\alpha}\left(\mathbf{r}\right)+v_{p}^{\mathrm{FOO}}\left(\mathbf{r}\right)$ in FOO
%%while there are $\left\lfloor N_{\alpha}\right\rfloor +1$ electrons subject to $v_{\alpha}\left(\mathbf{r}\right)+v_{p}^{\mathrm{ENS}}\left(\mathbf{r}\right)$ in ENS, $v_{p}^{\mathrm{ENS}}\left(\mathbf{r}\right)$ needs to be deeper than
%%$v_{p}^{\mathrm{FOO}}\left(\mathbf{r}\right)$ to account for the fact that
%%$\left\lfloor N_{\alpha}\right\rfloor +1>N_{\alpha}$. As a consequence,
%%$E_{p}^{\mathrm{ENS}}$ is more negative than $E_{p}^{\mathrm{FOO}}$ and, since FOO and ENS must yield the same total energy, $E_{f}^{\mathrm{ENS}}$ will be less negative 
%%than $E_{f}^{\mathrm{FOO}}$.   
%%All of these findings are illustrated below. 
%%%%%%%%
%  RESULTS %
%%%%%%%%
\section{Examples and Discussion}
% \label{sec:res}
%

% \newpage

We now illustrate the preceding discussions on a few model systems of diatomic molecules and on actual diatomic molecules. In all cases, the nuclei are separated a distance $R$ and located at $-(R/2)\hat{z}$ and $+(R/2)\hat{z}$ along the $z$-axis. 
%And we assign one nucleus to A fragment and the other nucleus to 
%B fragment.
All calculations are performed on a prolate-spheroidal real-space grid
\cite{becke_numerical_1982,nafziger_density-based_2014}.
%Kui: Why do we need so many references for a grid? 
Considering the azimuthal symmetry of diatomic molecules, 
we only need to solve the Kohn-Sham equations on
a two-dimensional mesh.
Libxc library \cite{marques_libxc:_2012} is used to evaluate 
approximate exchange correlation functionals.  
%The code employed for all calculations is available at: 
%Kui: Provide location of the code:  Please upload the code(s) onto our github repository (talk with Victor about how to do this).

We  partition the molecules into two fragments, labeled $A$ and $B$, and denote 
the {\em optimal} electron populations by 
$N_{A}^{*}$ and $N_{B}^{*}$.  When isolated ($R\to\infty$), the correct populations are denoted by 
$N_{A}^{0}$ and $N_{B}^{0}$, so we can define the number of electrons transferred from $A$ to $B$ as 
$\Delta N=N_{A}^{0}-N_{A}^{*}=N_{B}^{*}-N_{B}^{0}$.  When no superscript is used for $N_A$ and $N_B$, it should be understood that the numbers have been fixed, but not optimized (i.e. they do not minimize $E_f$).

\subsection{{\underline{One electron}}: Model of a one-electron molecule $\mathbf{AB^{+Z_{B}}}$ }

\begin{figure}[htp]
	\centering
	\includegraphics[width=1.0\linewidth]{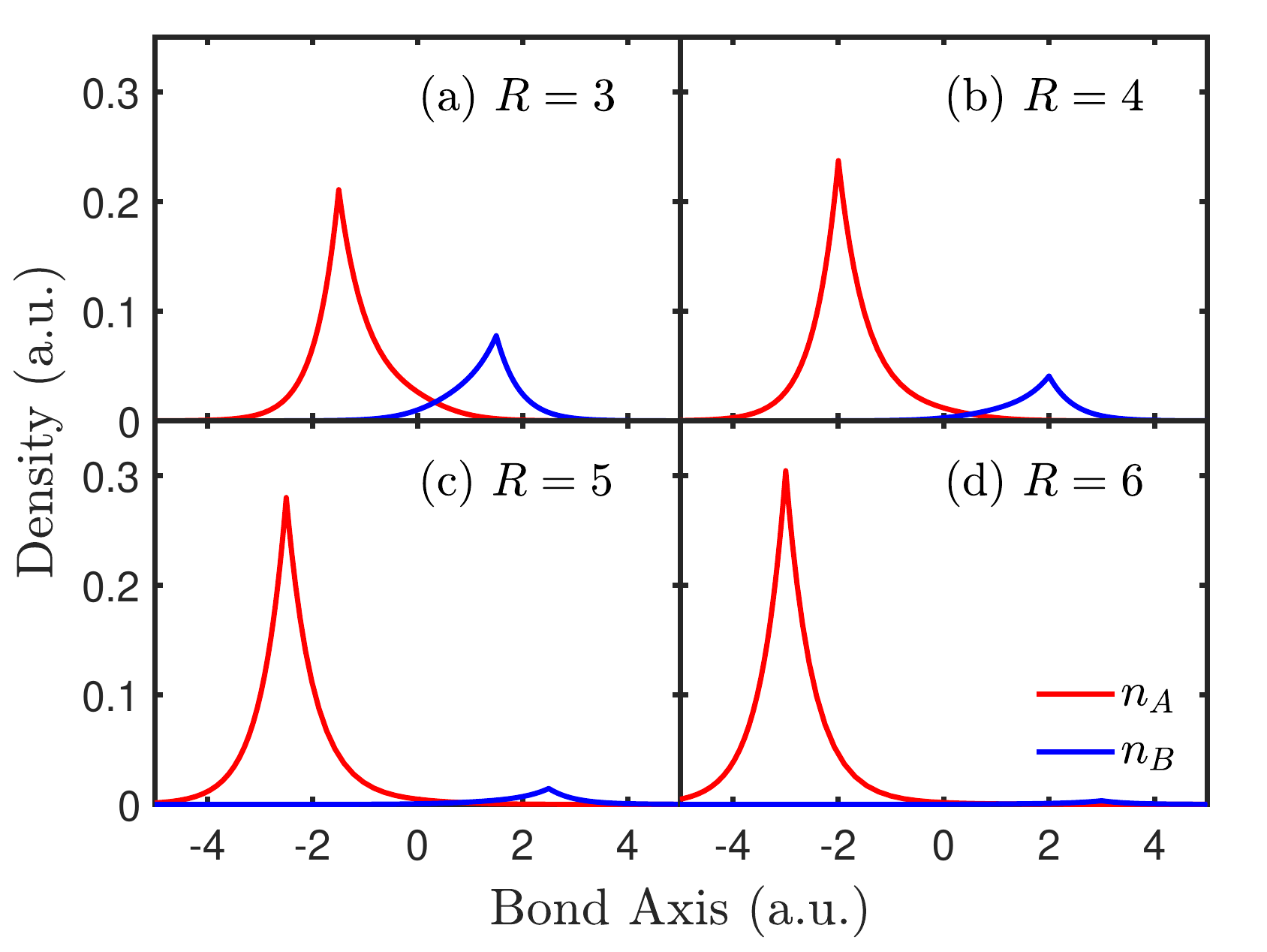}
	\caption{ The optimal electron densities of fragments along the bond axis for
	model molecule $\mathrm{AB}^{+0.9}$ ($Z_A=1$ and $Z_B=0.90$)
	with the exact functional
	when the internuclear distance is $R=$ (a) 3.0, (b) 4.0, 
	(c) 5.0, and (d) 6.0 a.u..}
	\label{fig:ABp_exact_f2}
\end{figure}

\begin{figure*}[htp]
	\centering
	\includegraphics[width=1.0\linewidth]{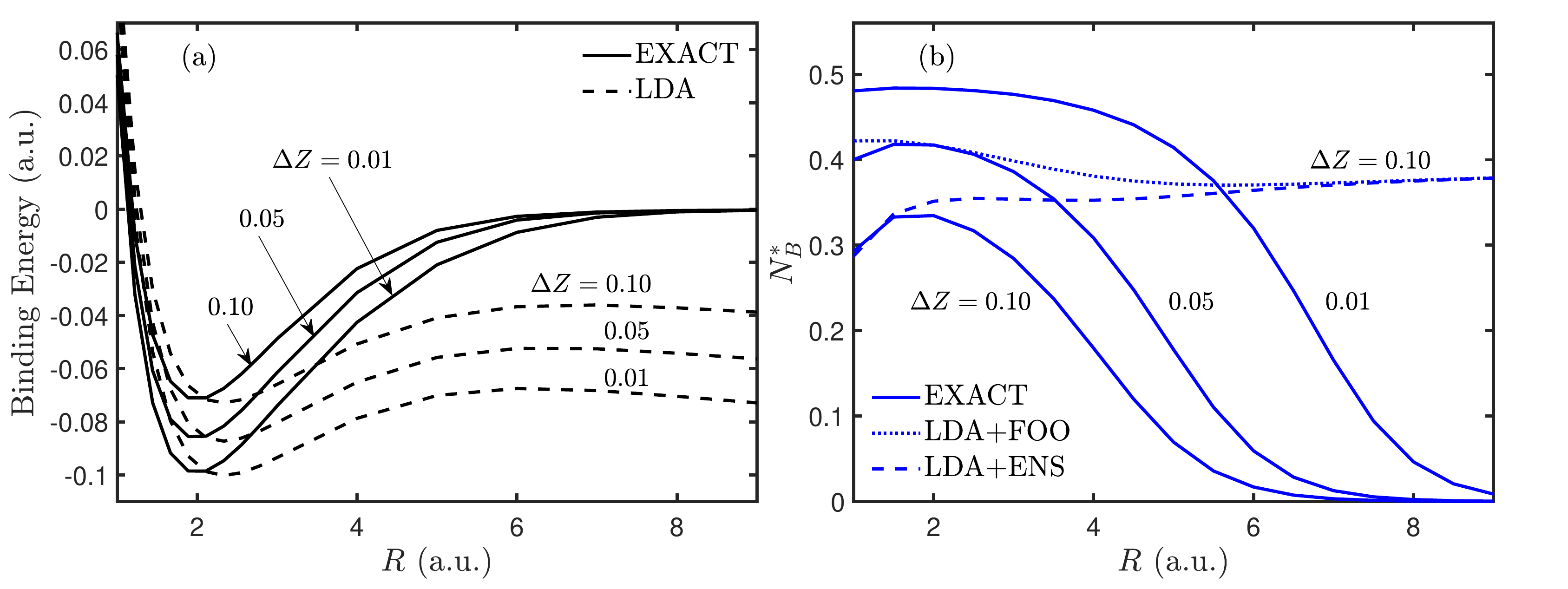}
	\caption{ Binding energies (left) and optimal population of fragment B (right) 
	for the model of a one-electron heteronuclear diatomic molecule $\mathrm{AB^{+Z_{B}}}$ (Sec.III-A) with the exact functional (solution) and LDA for $Z_A=1$, and $Z_B=0.90, 0.95, 0.99$. FOO and ENS results lead to identical binding energies, but $N_B^*$ differs when using the LDA as indicated in the right panel}.	
	\label{fig:ABp_disso}
\end{figure*}

We first consider a one-electron diatomic molecular ion $\mathrm{AB^{+Z_{B}}}$ with nuclear charges $Z_A=1$ and variable $Z_B<1$, so that the dissociated molecule consists of a hydrogen atom and a ``proton" of charge $+Z_B$. (In the discussion below, we sometimes refer to $\Delta Z\equiv 1-Z_B$).
%Since there is only one electron, it can be exactly calculated.
A similar model but in 1D and with delta-function potentials was studied in ref.\cite{cohen_charge_2009}. The {\em exact} P-DFT solution reported here for Coulomb potentials in 3D leads to well-localized fragment densities for all $R$ (see Figure \ref{fig:ABp_exact_f2}), just like in the 1D model of ref.\cite{cohen_charge_2009}. As the molecule is stretched, an electron is smoothly transferred from $B$ to $A$. (What we mean by ``smoothly" will be clarified below, see the right panel of Fig.\ref{fig:ABp_disso}).

The left panel of Fig.\ref{fig:ABp_disso} compares the {\em exact} and LDA dissociation curves of $\mathrm{AB^{+Z_{B}}}$, showing a vivid example of LDA fractional-charge and delocalization errors. The equilibrium distance $R_{\rm eq}$ is about 2 a.u. for the exact solution and about 2.3 a.u. for the LDA ($R_{\rm eq}$ is approximately constant in the range $0.9<Z_B<1$). Two lessons can be drawn from the left panel of Fig. \ref{fig:ABp_disso}: 

(1) The bond is stronger for the homonuclear case than it is for the heteronuclear case. The larger the value of $\Delta Z$, the weaker the binding, as explained beautifully in Chapter 10 of ref.\cite{feynman_feynman_2011}.
%Kui: This is Volume III of Feynman's Lectures.
%When $Z_{B}<1$, the molecule dissociates into one hydrogen atom and one ``proton" of charge is $Z_B$.
%(2) The LDA binding energies $E_b$ (defined as the difference between the energy of the molecule and the sum of the energies of the isolated fragments) do not vanish with increasing $R$, as they should. This error increases as $\Delta Z\to 0$. In the limiting case of $\Delta Z=0$ (i.e. H$_2^+$), the error has been well studied
%\cite{perdew_self-interaction_1981, cohen_insights_2008} and understood as due to the incorrect treatment of fractional charges by the LDA.  
%The Local Density Approximation (LDA) severely underestimates the dissociation energy of H$_2^+$. 

(2) The LDA error manifests most clearly at the dissociation limit as the binding energy $E_b$ goes incorrectly to a negative value ($E_b$ is defined as the difference between the ground-state energy of the molecule and the sum of the ground-state energies of the isolated atoms, in this case just a hydrogen atom, so should be zero at dissociation). Furthermore, Figure \ref{fig:ABp_disso} shows that this error increases as $\Delta Z\to 0$. In the limiting case of $\Delta Z=0$ (i.e. H$_2^+$), the error has been well studied
\cite{perdew_self-interaction_1981, cohen_insights_2008} and understood as due to the incorrect treatment of fractional charges by the LDA.  
%this negative value changes with $\Delta Z$ (the difference between the two nuclear charges), which is zero for H$_2^+$ but can be tuned artificially to model a one-electron heteronuclear diatomic molecule. While the exact binding energy goes to zero as the inter-nuclear separation $R$ grows, regardless of $\Delta Z$, the LDA binding energy goes to a negative constant that depends strongly on $\Delta Z$. 
A hallmark of the delocalization error (DE) for a heteronuclear diatomic molecule is that the approximate functional incorrectly minimizes the total energy by placing fractional charges on the separated atoms. These incorrect fractional charges are well defined at dissociation. An interesting question arises here: The (incorrect) fractional charges that determine the DE are strictly only well defined at dissociation, as mentioned in the Introduction, but the left panel of Figure \ref{fig:ABp_disso} shows that the DE settles in slowly as $R\gtrsim 4$. How do the fractional charges at {\em finite} $R$ evolve into those at dissociation as $R\to\infty$?  
%The right panel of Figure 1 also shows that the LDA error settles in for finite $R$ way before reaching dissociation, where the concept of ``atomic charge" cannot be sharply defined. 
%``How many electrons belong to a fragment in a molecule?" is an ill-posed question, as there is no unique way to define a fragment in a molecule. The number of electrons in a molecular fragment is not an observable, except at dissociation. Nevertheless, 

The right panel of Fig.\ref{fig:ABp_disso} provides the answer given by Partition-DFT through both ENS and FOO methods described in Sec.II, using both the exact functional ($E_{\mathrm{XC}}^{\mathrm{EXACT}}[n]=-E_{\mathrm{H}}[n]$ for one electron) and the LDA.  For the exact case, ENS and FOO fragments are identical. The number of electrons in the $B$-fragment ($N_B$) reaches a maximum value for small $R$ and decreases monotonically down to zero at dissociation in the same way as was observed for the 1D model system of ref.\cite{cohen_charge_2009}.  However, in the case of an approximate functional like the LDA, though FOO and ENS yield the same {\em total} molecular density and energy, 
they yield different fragment energies and densities. The ENS-LDA $N_B$ values are close to the exact ones for very small internuclear separations and around the maximum of $N_B$, but they do not decrease as $R$ grows and stay approximately constant for $R\gtrsim 4$, suggesting that the error in $N_B$ encodes the error in $E_b$. The FOO-LDA $N_B$ values converge to those of ENS-LDA as $R$ grows, but differ significantly for small $R$. 

To discuss the origin of these differences, we take the case of $Z_{B}=0.9$ as an example, and fix $R=2.0$ a.u. (close to equilibrium). 

%For a diatomic molecule, we always can divide it into two fragments.

% \subsubsection{LDA}

\begin{figure*}[htp]
\centering
  \includegraphics[width=1.0\linewidth]{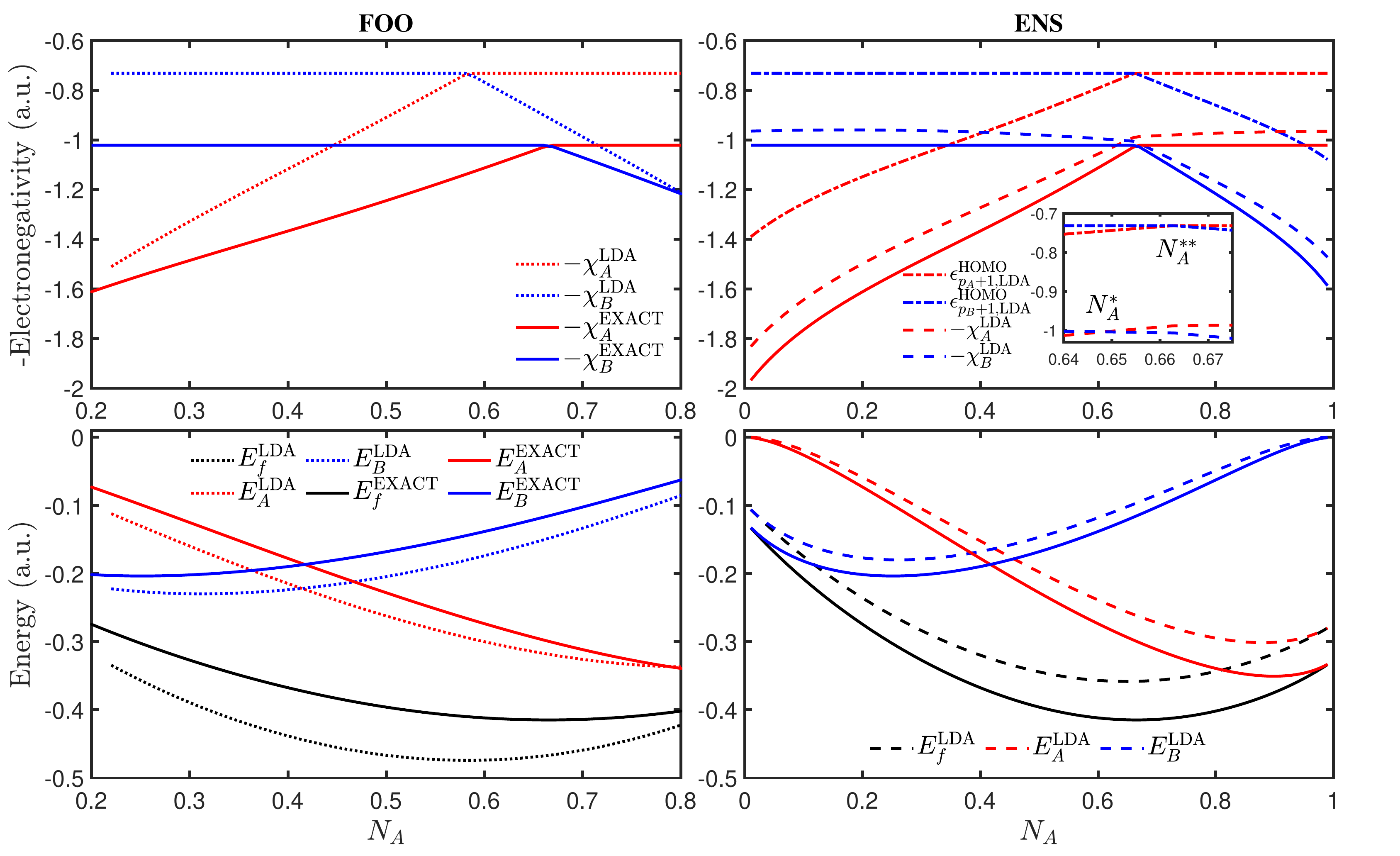}
  %Kui: Please stick a "minus" sign before "Electronegativity" on the Y-axis of the top-left panel.
  \caption{Negative electronegativities of fragments (top panels) and fragment          energies and their sum (bottom panels) as a function of $N_A$ for 
    the one-electron model molecule $\mathrm{AB^{+0.9}}$ of Sec.III-A ($Z_A=1$ and $Z_B=0.9$; at internuclear distance $R=2.0$ a.u.. $\epsilon^{\mathrm{HOMO}}_{M,\mathrm{LDA}}=-0.7311$ a.u. and
    $\epsilon^{\mathrm{HOMO}}_{M,\mathrm{EXACT}}=-1.0216$ a.u.) with LDA and exact functionals. Left panels: FOO. Right panels: ENS. 
  }
\label{fig:ABp_lda_f1}
\end{figure*}

{\underline{\em FOO:}} The LDA electronegativities $\chi_A$ and $\chi_B$, as defined below Eq.(\ref{total_frag_energy}), become equal when $N_A^*=0.5830$ (for the exact case, they equalize at $N_A^*=0.6654$, so the LDA error for the optimum occupation is -12\%). This number is also the minimizer of $E_f$ (see bottom panels of Fig.\ref{fig:ABp_lda_f1}). When $N_A<N_A^*$, the electronegativity of $B$ equals $-\epsilon_{M}^{\mathrm{HOMO}}$, a constant value. In this range, $\chi_A>\chi_B$, so fragment $A$ has a higher tendency than $B$ to attract electrons: $B$ is the donor (``base") and $A$ is the acceptor (``acid"). As electrons are transferred from $B$ to $A$, both the electronegativity difference $\chi_{A}-\chi_{B}$ and $E_f$ decrease.  On the other hand, when $N_A>N_A^*$, fragment $A$ and $B$ swap their donor/acceptor roles.

%The left panels in Figure \ref{fig:ABp_lda_f1} show results of FOO 
%with LDA and exact functional.
%For LDA, $N_{A}^{*}=0.5830$ since $\chi_{A}=\chi_{B}$ and $E_f$ reaches its global minimum at this point.
%When $N_{A}<0.5830$, $\chi_{B}$ equals $-\epsilon_{\mathrm{HOMO}}^{M}=0.7311$ a.u. 
%that is a constant. And $\chi_{A}$ is bigger than $\chi_{B}$, 
%which means fragment A tends to attract more electrons than fragment B. 
%Therefore, if electrons are transferred from fragment B to A, 
%both the electronegativity difference $\chi_{A}-\chi_{B}$ and $E_f$ decrease. 
%And fragment A and B swap their roles when $N_{A}>0.5830$.
%For exact functional, $N_{A}^{*}=0.6654$ which is bigger than 0.5830.

{\underline{\em ENS:}} The LDA electronegativities now cross at $N_A^*=0.6485$. By switching from FOO to ENS, the magnitude of the error in $N_A^*$ has thus been reduced from 12\% to 2\%.   
%The right panels in Figure \ref{fig:ABp_lda_f1} show results of ENS 
%with LDA and exact functional.
%For LDA, $N_{A}^{*}=0.6485$ since $\chi_{A}=\chi_{B}$ and $E_f$ reaches 
%its global minimum at this point.
%When $N_{A}<0.6485$, $\chi_{A}>\chi_{B}$, if electrons are transferred 
%from fragmemnt B to A, 
%both $\chi_{A}-\chi_{B}$ and $E_f$ decrease. 
%And fragment A and B swap their roles when $N_{A}>0.6485$.
%Negative electronegativities of fragments are continuous 
%when $N_A$ is non-integral.
%And there always exist one fragment whose HOMO energy is equal to 
The HOMO energy of the molecule equals the HOMO energy of one of the two fragments. That fragment is $A$ when $N_A>N_A^{**}$ and $B$ when $N_A<N_A^{**}$, where $N_A^{**}$ is the crossing point of the fragment HOMO energies, which differs slightly from $N_A^*$ (the crossing point for the electronegativities). In the case of Fig.{\ref{fig:ABp_lda_f1}} (see inset in top right panel), $N_A^{**}=0.6630$. 
%Kui: Add the value of N_A^**
%.  In agreement with our discussion in Sec.II-C, we note that the value of $N_A$ at which 
%We also notice that the cross point of 
%$\epsilon_{A}^{\mathrm{HOMO}}$ becomes equal to 
%$\epsilon_{B}^{\mathrm{HOMO}}$ is slightly different from $N_A^*$. (see the inset in the top right panel of Fig.(\ref{fig:ABp_lda_f1})). 
%$N_{A}^{*}=0.6485$ is also less than 0.6645 which is the $N_{A}^{*}$ 
%for exact functional.

This example indicates that the FOO method leads to a larger estimate than ENS for the charge transferred. In the case of Fig.(\ref{fig:ABp_lda_f1}),  
%By using the optimal electron populations, 
%we can obtain the electrons transferred from A to B:
$\Delta  N^{\mathrm{FOO}}=0.4170$ and $\Delta N^{\mathrm{ENS}}=0.3515$. 
%Therefore, the number of electrons transferred in FOO is larger than that in ENS.

\begin{figure}[htp]
	\centering
	\includegraphics[width=1.0\linewidth]{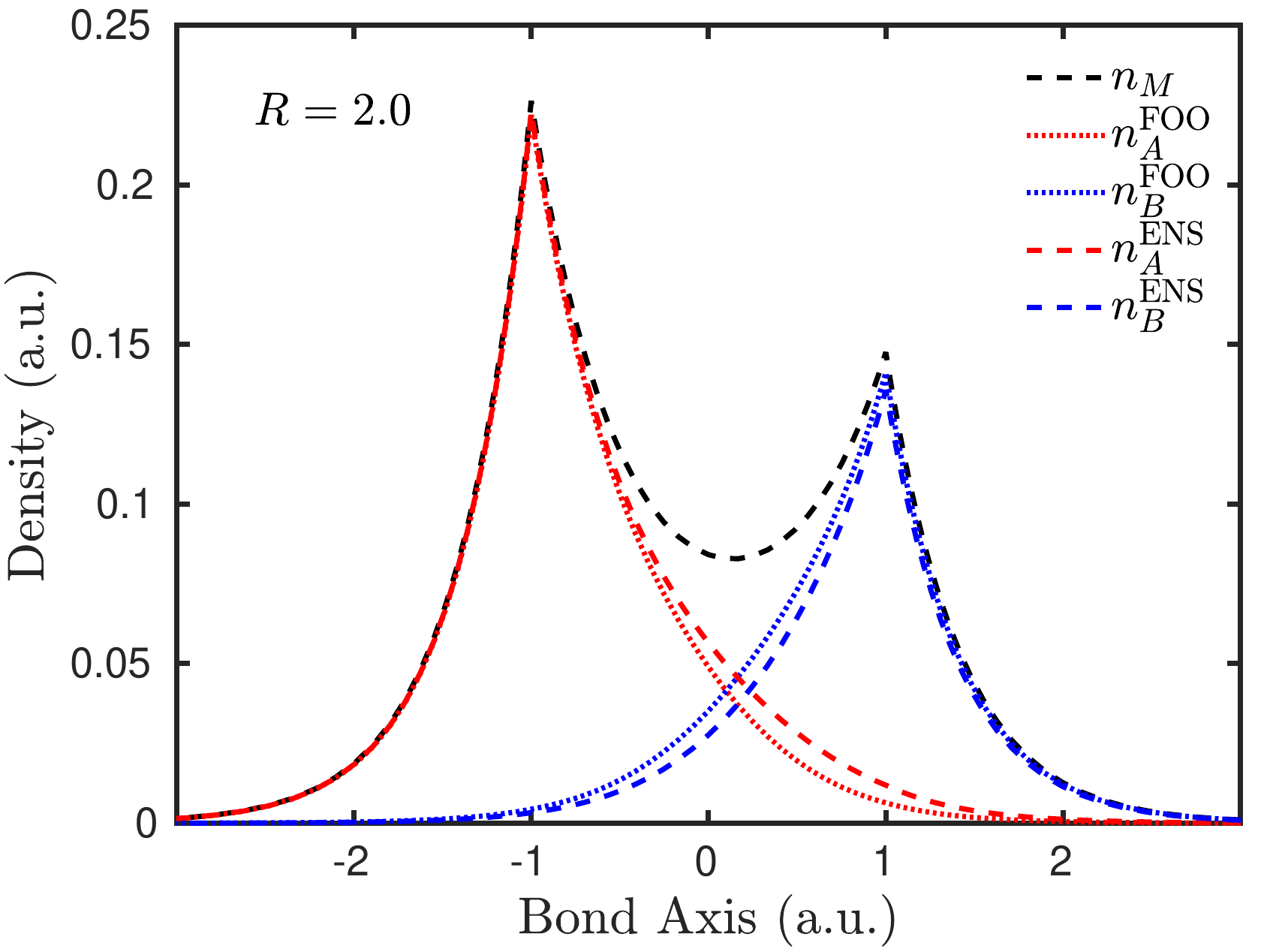}
	\caption{ Fragment densities of the one-electron model molecule $\mathrm{AB^{+0.9}}$ ($Z_A=1$ and $Z_B=0.9$) along the bond axis
	at optimal electron populations with FOO+LDA and ENS+LDA 
	 when the internuclear distance is $R=2.0$ a.u..}	
	\label{fig:ABp_lda_f5}
\end{figure}

Now we note another qualitative difference between the two methods for calculating fractional populations: The ENS-LDA values of $E_f$ are above the exact ones, but the FOO-LDA values are below.  This behavior can be traced back to an increased delocalization of the LDA-FOO densities when compared to ENS (see Figure \ref{fig:ABp_lda_f5}), which in turn leads to an increased magnitude for the (negative) electron-nuclear interaction energy in FOO. 
This observation agrees with our analysis in Sec.II-C.
%Kui:  You said you wanted more explanations here. Please e-mail me a paragraph labeled "A" 

%Figure \ref{fig:ABp_lda_f5} shows fragment densities along the bond axis 
%for FOO and ENS with LDA. 
% FOO and ENS treatments divide the same total density into different fragment densities.
%Related to this observation, the FOO fragment dipole moments
Setting the bond mid-point as the origin, we define 
fragment electronic dipole moments as 
$\mathbf{p}_{\alpha}=\int\mathbf{r}n_{\alpha}\left(\mathbf{r}\right)d\mathbf{r}$.
Then,
$\mathbf{p}_{A}^{\mathrm{FOO}}=0.4219>0.3962=\mathbf{p}_{A}^{\mathrm{ENS}}$
and 
$\left|\mathbf{p}_{B}^{\mathrm{FOO}}\right|=\left|-0.2942\right|>\left|-0.2685\right|=\left|\mathbf{p}_{B}^{\mathrm{ENS}}\right|$. Clearly, the FOO fragments are more polarized than the ENS fragments. 

\subsection{{\underline {Two electrons:}} Model of a two-electron molecule $\mathbf{AB^{+\delta}}$ }

We now add one more electron and allow both electrons to fully interact with each other and with the two nuclei. To keep the system bound in the LDA, this time we fix $Z_B=1$ and let $Z_{A}\in\left[1,2\right]$, so we vary $
\delta\equiv Z_A-Z_B$ from zero  (H$_2$) to one (HeH$^+$). 
While the fragment electron populations were always non-integral for the previous one-electron diatomic molecule, we will show here that, for a range of $\delta$, the electron-electron interaction can make the fragments acquire strictly integer populations,
%However, for a two-electron diatomic molecule, the electron populations of 
%fragments can be integers, 
i.e. $N_{A}^*=N_{B}^*=1$.
%Now we consider a model of a diatomic molecule with two electrons.
%We set $Z_{B}=1$, and $Z_{A}\in\left[1,2\right]$. 
%When $Z_{A}=1$, it is a hydrogen molecule.
%When $Z_{A}=2$, it is a helium hydride ion.

We consider two separation channels in the LDA:
{\underline {\em Channel $\mathrm{\uppercase\expandafter{\romannumeral1}}$}}: 
%And the separated state can be determined by comparing the total energies of them.
%We use KS-DFT with LDA to calculate their energies.
When $1\leq Z_{A}<1.607$, the separated state is one hydrogen atom 
and a one-electron atom whose nuclear charge is $Z_{A}$.
{\underline {\em Channel $\mathrm{\uppercase\expandafter{\romannumeral2}}$}}:
When $1.607<Z_{A}\leq 2$, the separated state is one proton and a two-electron 
atom with nuclear charge $Z_{A}$.

\begin{figure*}[htp]
\centering
  \includegraphics[width=1.0\linewidth]{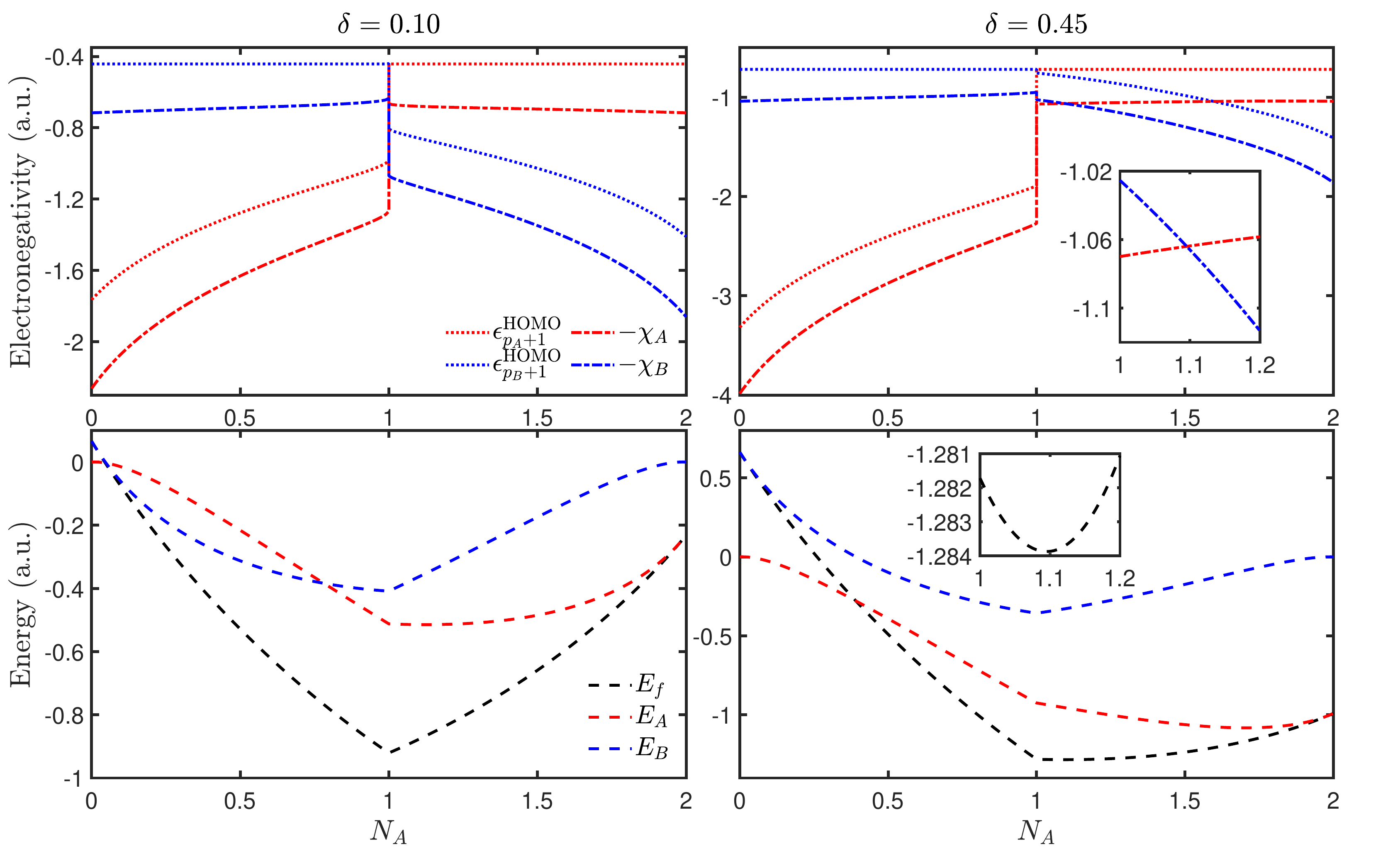}
  \caption{Top left: Negative electronegativities of fragments and HOMO energies as a function of $N_A$ for the two-electron molecule 
  $\mathrm{AB^{+\delta}}$ of Sec.III-B ($Z_A=1.1$ and $Z_B=1$) with LDA and ENS 
  at internuclear distance $R=1.446$ a.u.. 
  $\epsilon_{M}^{\mathrm{HOMO}}=-0.4419$ a.u..
  Bottom left: Fragment energies and their sum as a function of $N_A$.
  Top and bottom right panels are the same as top and bottom left, respectively, but with $Z_A=1.45$.
  }
\label{fig:AB_fig1}
\end{figure*}

\begin{figure}[htp]
	\centering
	\includegraphics[width=1.0\linewidth]{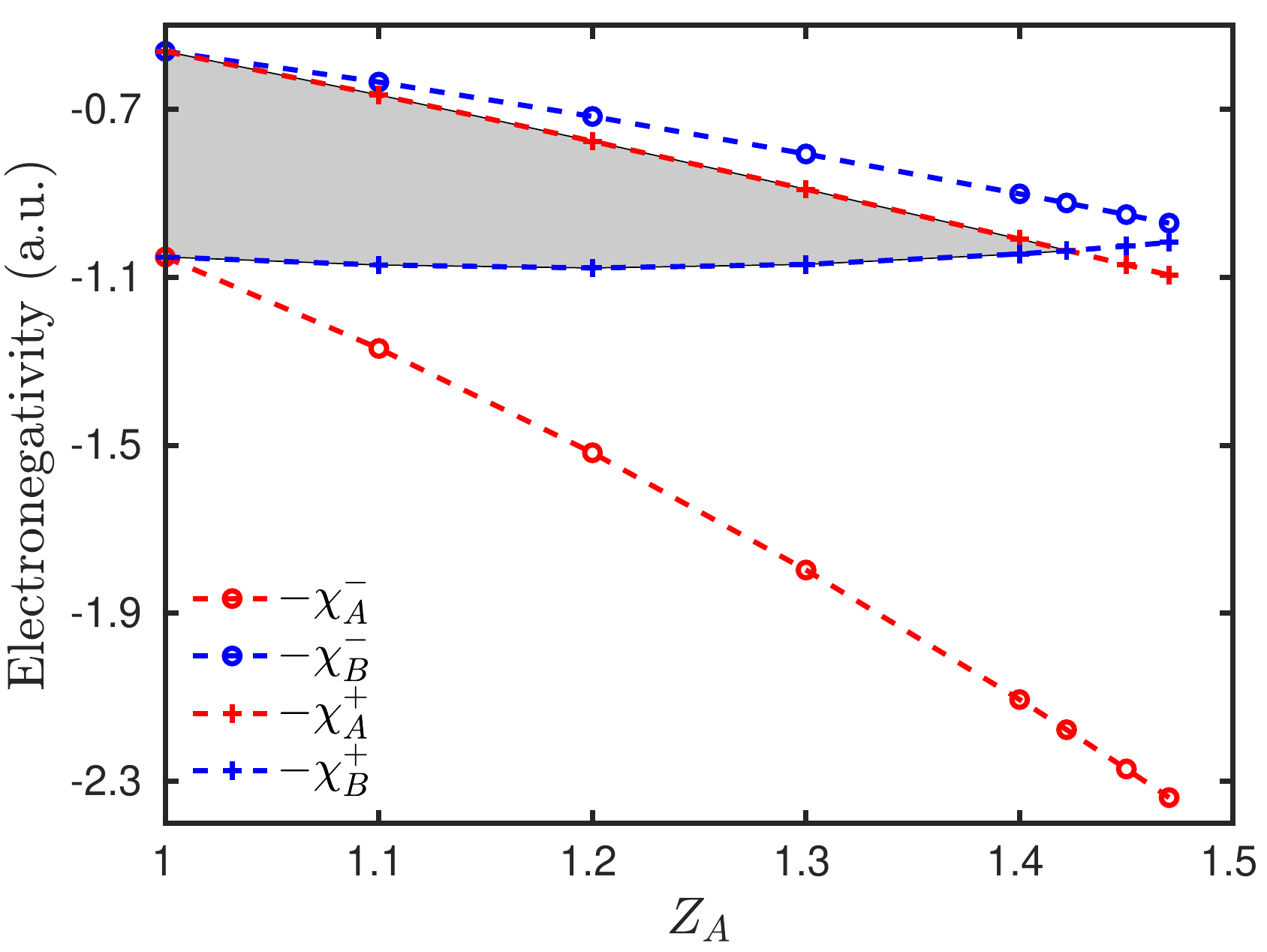}
	\caption{ENS-LDA left and right electronegativities of fragment A and B as a function of $Z_A$ for the two-electron model molecule $\mathrm{AB^{+\delta}}$ of Sec.III-B ($Z_B=1$) at internuclear distance $R=1.446$ a.u. when $N_A=N_B=1$. The shaded area corresponds to the range of $Z_A$ for which the optimum fragment populations are strictly integers.}
	\label{fig:en_limit}
\end{figure}

We begin by finding the optimum populations by identifying the value of $N_A$ at which the electronegativities of $A$ and $B$ become equal. Special care is needed here, however,  
as the ENS electronegativities are undefined at strictly integer populations, but one can define left- (right-) electronegativities, $\chi^{-}_{\alpha}$ ($\chi^{+}_{\alpha}$), 
as the left (right) limits
of $-\frac{\partial\widetilde{E_{\alpha}}\left[n_{\alpha}\right]}{\partial n_{\alpha}\left(\mathbf{r}\right)}$.
Figure \ref{fig:AB_fig1} shows that 
fragment electronegativities are discontinuous at $N_{A}=1$.
When $Z_{A}=1.1$, $N_{A}=1$ is the global minimizer of $E_{f}$.
However, when $Z_{A}=1.45$, $N_{A}=1$ is no longer the global minimizer (the inset in the lower right panel of Fig.\ref{fig:AB_fig1} shows that $N_A^*=1.1$ leads to a slightly lower $E_f$ than $N_A=1.0$). 
%If fragment electronegativities are well-defined in P-DFT, 
%$E_{f}$ reaches its global minimum when all fragment electronegativities are equal.
%Now the fragment electronegativities are not well-defined.
How can one know if the global minimizer 
of $E_f$ will involve integer or fractional populations?
Figure \ref{fig:en_limit} provides the answer.  The left and right electronegativites at 
$N_A=1$ are shown here as a function of $Z_A$. 
We find that 
when $1\leq Z_{A}\leq 1.422$, 
the intersection of
$\left[-\chi_{A}^{-},-\chi_{A}^{+}\right]$
and 
$\left[-\chi_{B}^{+},-\chi_{B}^{-}\right]$ 
is not empty (shaded area in Fig.\ref{fig:en_limit}), 
and $N_A=1$ is the global minimizer of $E_f$.
Otherwise, 
$N_A=1$ is no longer the global minimizer of $E_f$, and the optimum populations become fractional, in 
agreement with the electronegativity equalization principle.
% Kui:  You said you want to add something here about Parr's electronegativity. Please e-mail me a paragraph labaled "B".

The FOO electronegativities are
well defined when the electron populations are integers \cite{janak_proof_1978}. There is no need to define left and right quantities here, as in ENS. The behavior of the $\chi_{\alpha}^{\rm FOO}$'s is similar to that of the one-electron electronegativites of Fig.(\ref{fig:ABp_lda_f1}).

\begin{figure}[htp]
	\centering
	\includegraphics[width=1.0\linewidth]{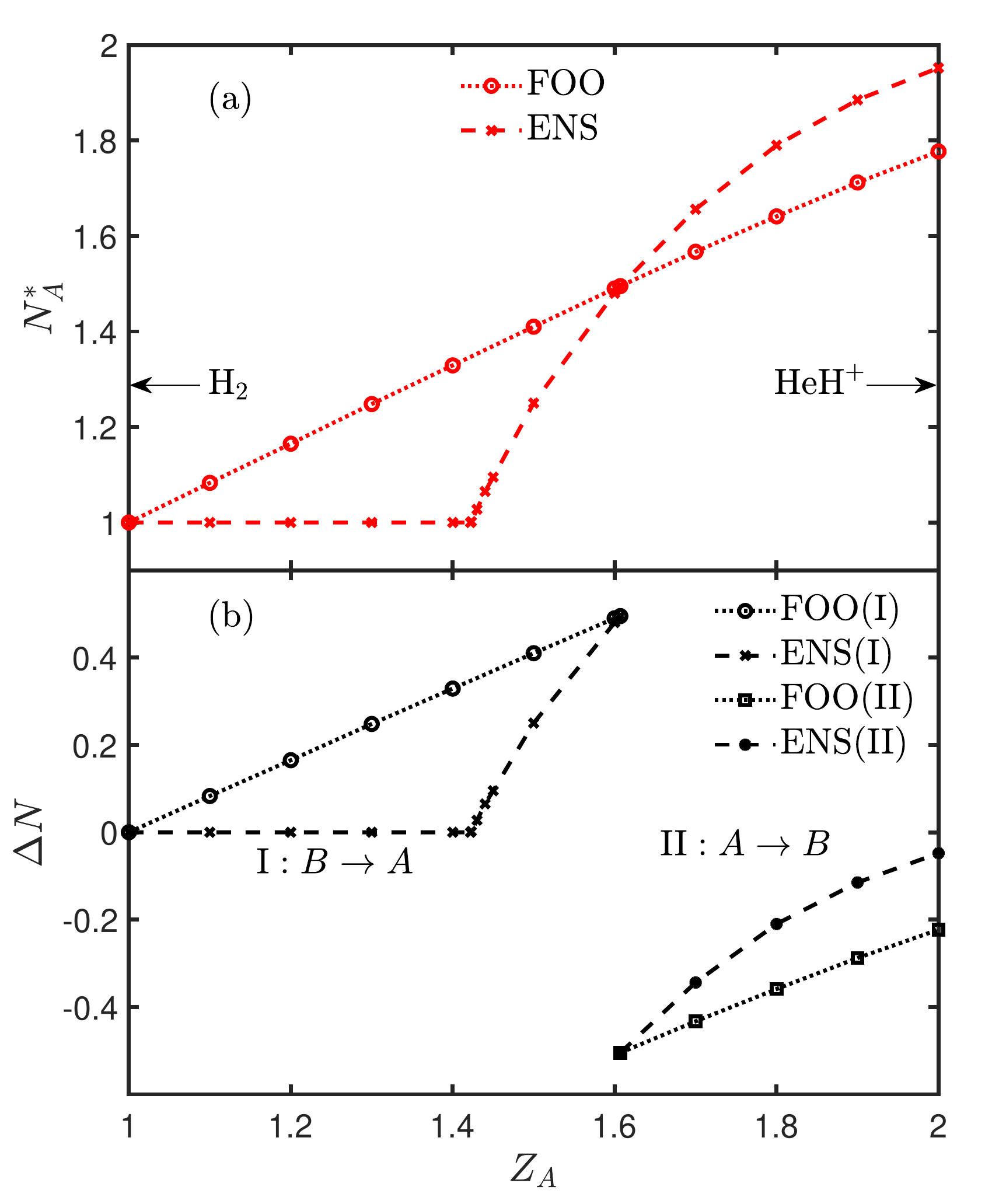}
	\caption{ {\em Top panel}: Optimal electron population of fragment A, $N_A^*$, as a function of 
	$Z_A$ for the two-electron model molecule $\mathrm{AB^{+\delta}}$ ($Z_B=1$) with FOO and ENS, at internuclear distance $R=1.446$ a.u. {\em Bottom panel}:
	The number of electrons transferred from fragment B to A, $\Delta N$, as a function of $Z_{A}$
	for the two dissociation channels discussed in the text.}
	\label{fig:AB_NA}
\end{figure}

\begin{figure}[htp]
	\centering
	\includegraphics[width=1.0\linewidth]{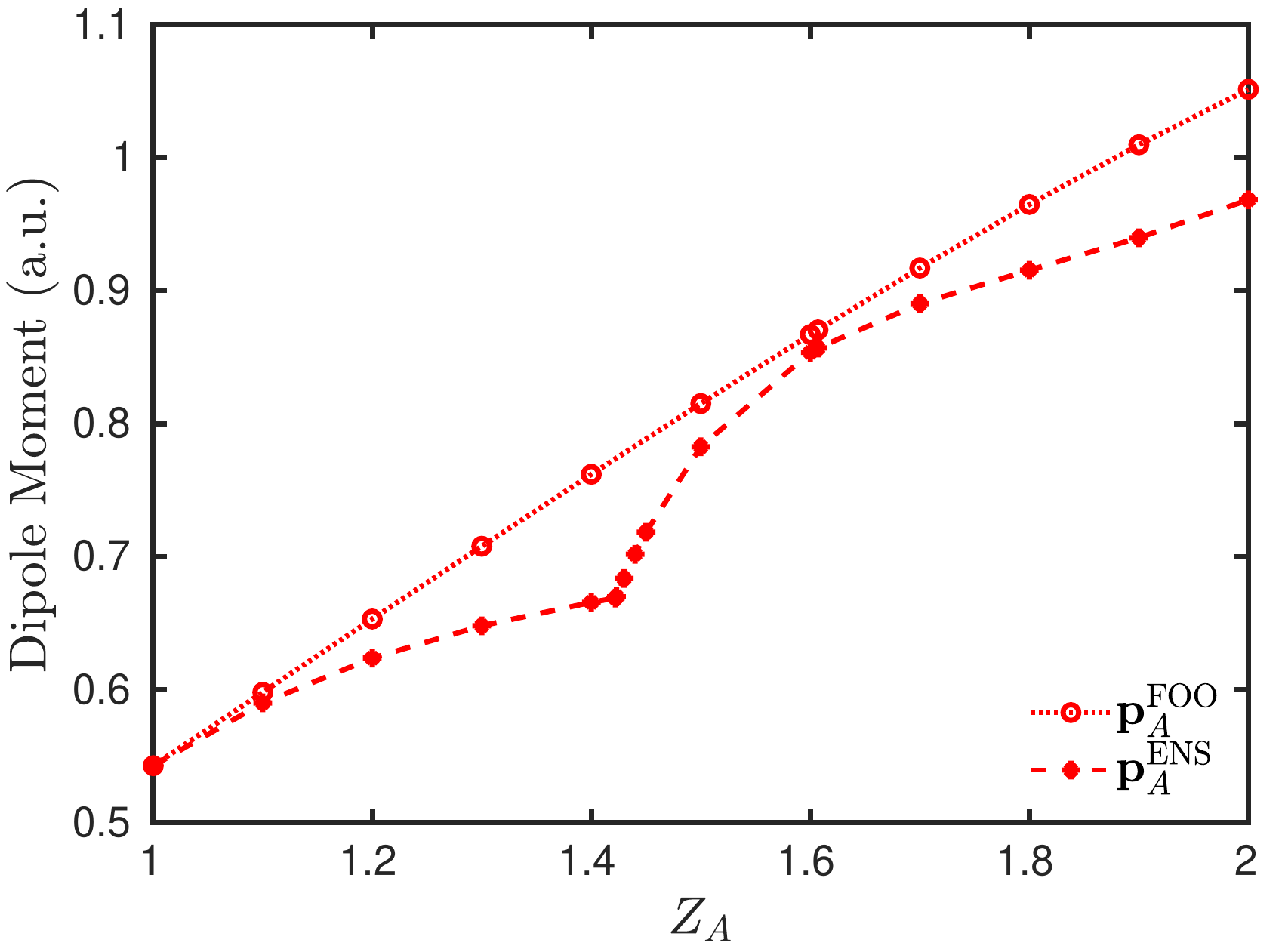}
	\caption{Electronic dipole moments of fragment A as a function of 
	$Z_A$ for the two-electron model molecule $\mathrm{AB^{+\delta}}$ ($Z_B=1$)
	with FOO and ENS. }
	\label{fig:AB_dipole}
\end{figure}

Figure \ref{fig:AB_NA} compares $N_A^*(Z_A)$ from FOO and ENS. 
While the optimum numbers are non-integers in FOO for the entire range $1< Z_{A} \leq 2$, they are integers in ENS when $1< Z_{A}\leq 1.422$ for the reasons mentioned above.
%So even for heteronuclear diatomic molecules, ENS treatment can give 
%integer electron populations.
Figure \ref{fig:AB_NA} also shows that FOO and ENS yield the same $N_{A}^*$ at $Z_{A}=1.607$. This is the critical nuclear charge above which Channel II becomes the ground state. To emphasize this point, Figure \ref{fig:AB_NA} shows the number of electrons transferred $\Delta N$ for the range $1<Z_A<2$, indicating the ground-state dissociation channel in each case. Electrons are transferred from $B$ to $A$ when $Z_A < 1.607$ and from $A$ to $B$ when $Z_A>1.607$. We note that $|\Delta N|$ is always lower in ENS than in FOO. The ENS fragment dipoles are also always smaller (Fig.\ref{fig:AB_dipole}) .

\begin{figure}[htp]
	\centering
	\includegraphics[width=1.0\linewidth]{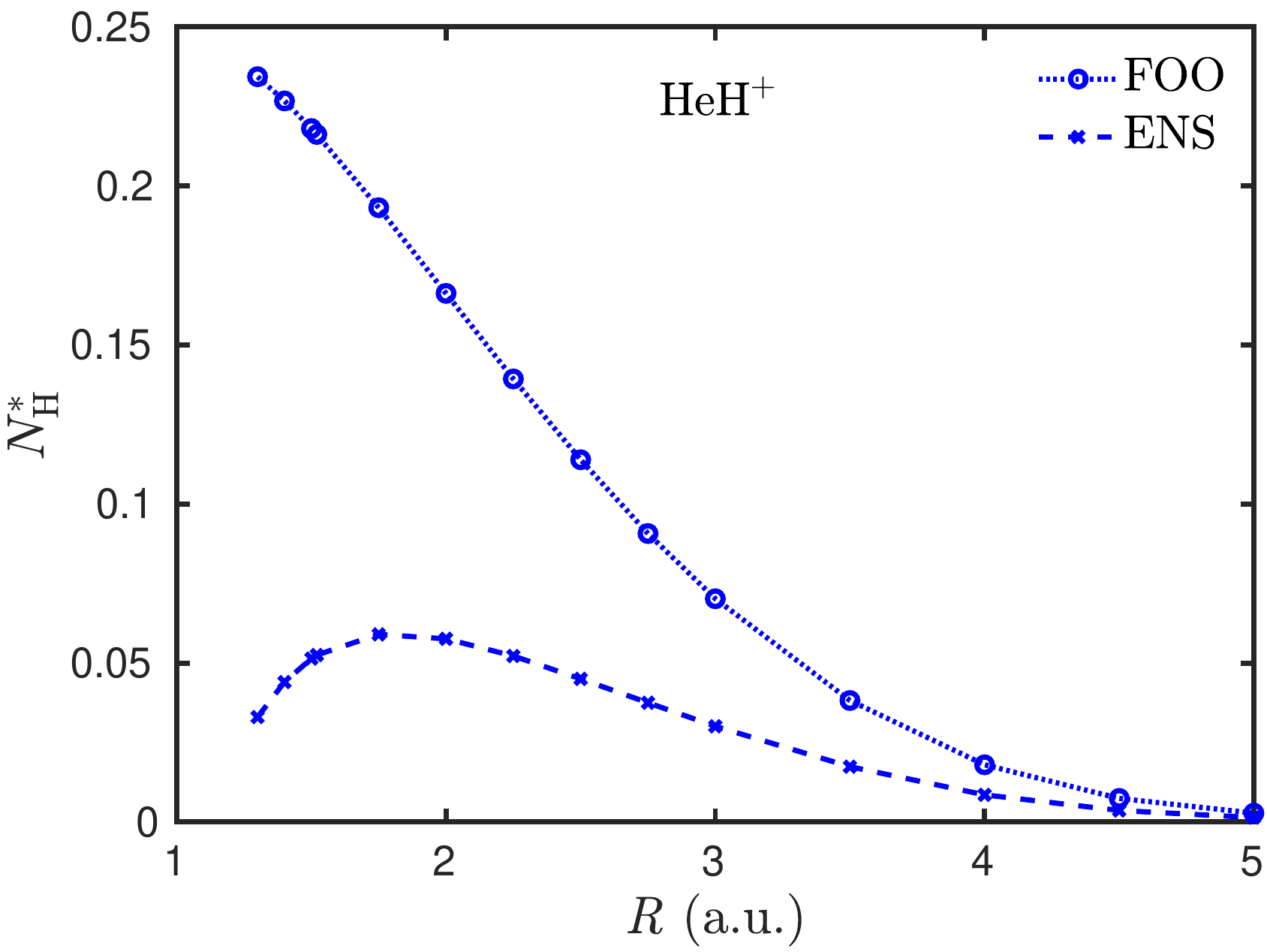}
	\caption{Optimal electron population of fragment H, $N_{\rm H}^*$, as a function of 
	internuclear distance for $\mathrm{HeH}^{+}$ with FOO and ENS.}
	\label{fig:diss_HeHp}
\end{figure}

{\underline{Dissociation of HeH${^+}$:}} 
We now compare FOO and ENS results for the optimum number of electrons transferred in HeH$^+$ as the internuclear separation increases from $R\sim 1$ a.u. up to $R\sim 5$ a.u. (Figure \ref{fig:diss_HeHp}). 
The separated state of $\mathrm{HeH}^{+}$ is a helium atom and a proton. The proton is dressed with a small fraction of an electron at finite $R$ and 
we note that the qualitative behavior of this small fraction $N_{\rm H}^*(R)$ is similar to that observed in the exact case for the 1-electron molecule of Sec.III-A (Figure \ref{fig:ABp_disso}): (1) The fraction of an electron reaches a maximum for ENS at $R\sim 1.8$ a.u.;
%Kui: Is the maximum around 1.8? 
(2) The FOO and ENS values approach each other as $R$ grows; and (3) 
The electron fraction given by FOO is larger than that given by ENS for all $R$.

% When the internuclear distance is large and we use LDA, 
% the binding energy does not go to zero but increases above zero.
% Therefore, we only consider the equilibrium distance. 

% For FOO, when the internuclear distance increases, 
% $N_{\mathrm{H}}$ does not go to one 
% since LDA gives the wrong binding curve.

% Q1: For exact functional, does $N_{\mathrm{H}}$ go to one for large 
% internuclear distance?

% For ENS, $N_{\mathrm{He}}=3$ and $N_{\mathrm{H}}=1$ for different 
% internuclear distances, thus, the charges transferred are zero.
% However, the centers of charge are not the position of nuclei.
% The distribution of the electrons are distorted.

% Q1: For the exact functional, does $N_{\mathrm{H}}$ still one for different 
% internuclear distance?

% Q2: For other molecules, are electron populations integral for different 
% internuclear distances?

% dipole moment

% Figure 3: components of $E_f$

% We should answer the question: why Foo gives bigger charges transferred than ENS?

% \newpage

\subsection{{\underline{Four electrons}}: Lithium hydride} 

%Kui:  You wanted more explanations here for v_p^ENS vs. v_p^FOO. Please add these brief explanations in relation to the figure of the partition potentials and e-mail me a paragraph labeled "C".

\begin{figure}[htp]
	\centering
	\includegraphics[width=1.0\linewidth]{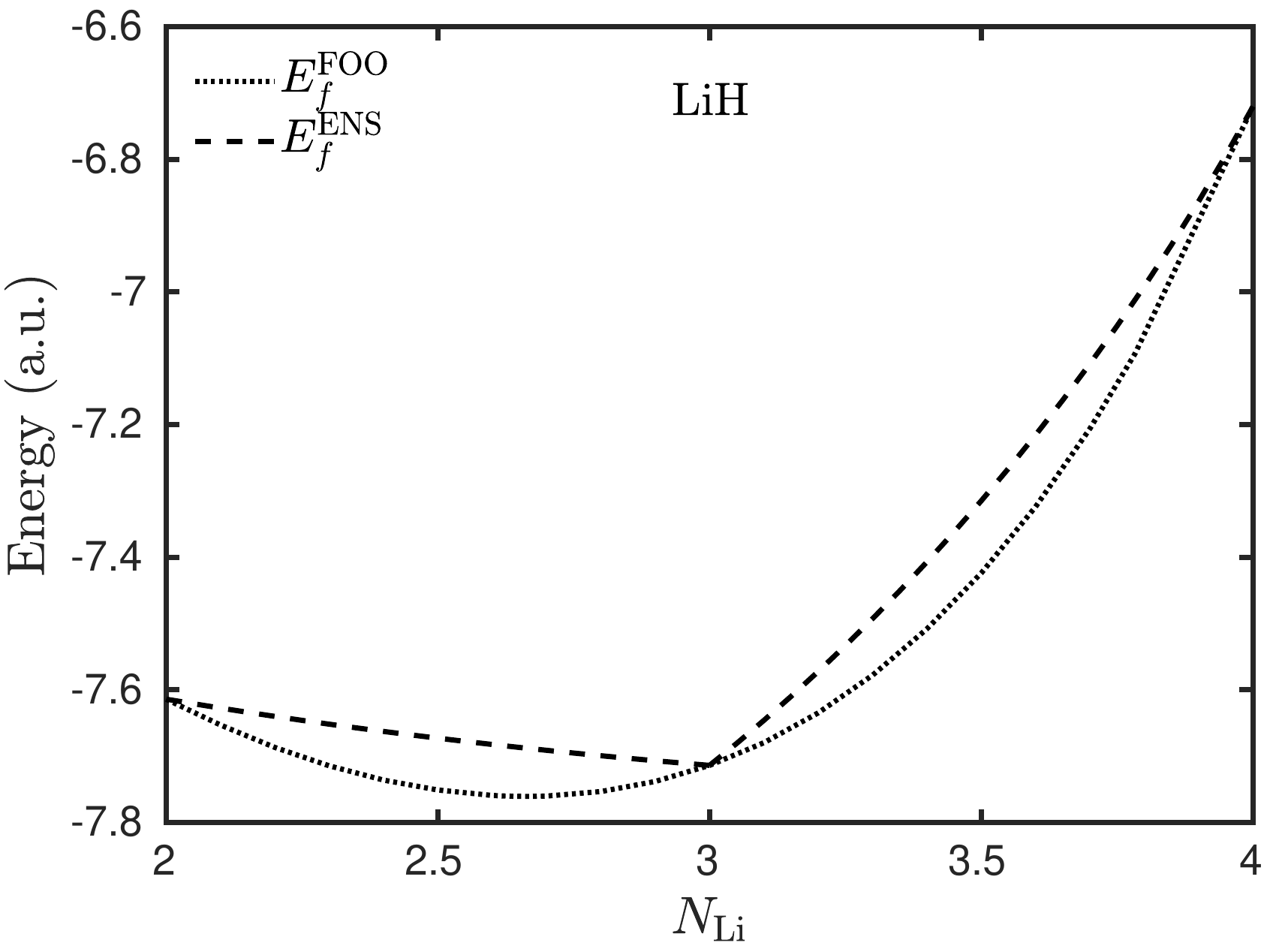}
	\caption{ The sum of fragment energies, $E_{f}$, as a function of $N_{A}$ 
	for $\mathrm{LiH}$ with FOO and ENS at the equilibrium internuclear distance $R=3.03$ a.u. }
	\label{fig:LiH}
\end{figure}

\begin{figure}[htp]
	\centering
	\includegraphics[width=1.0\linewidth]{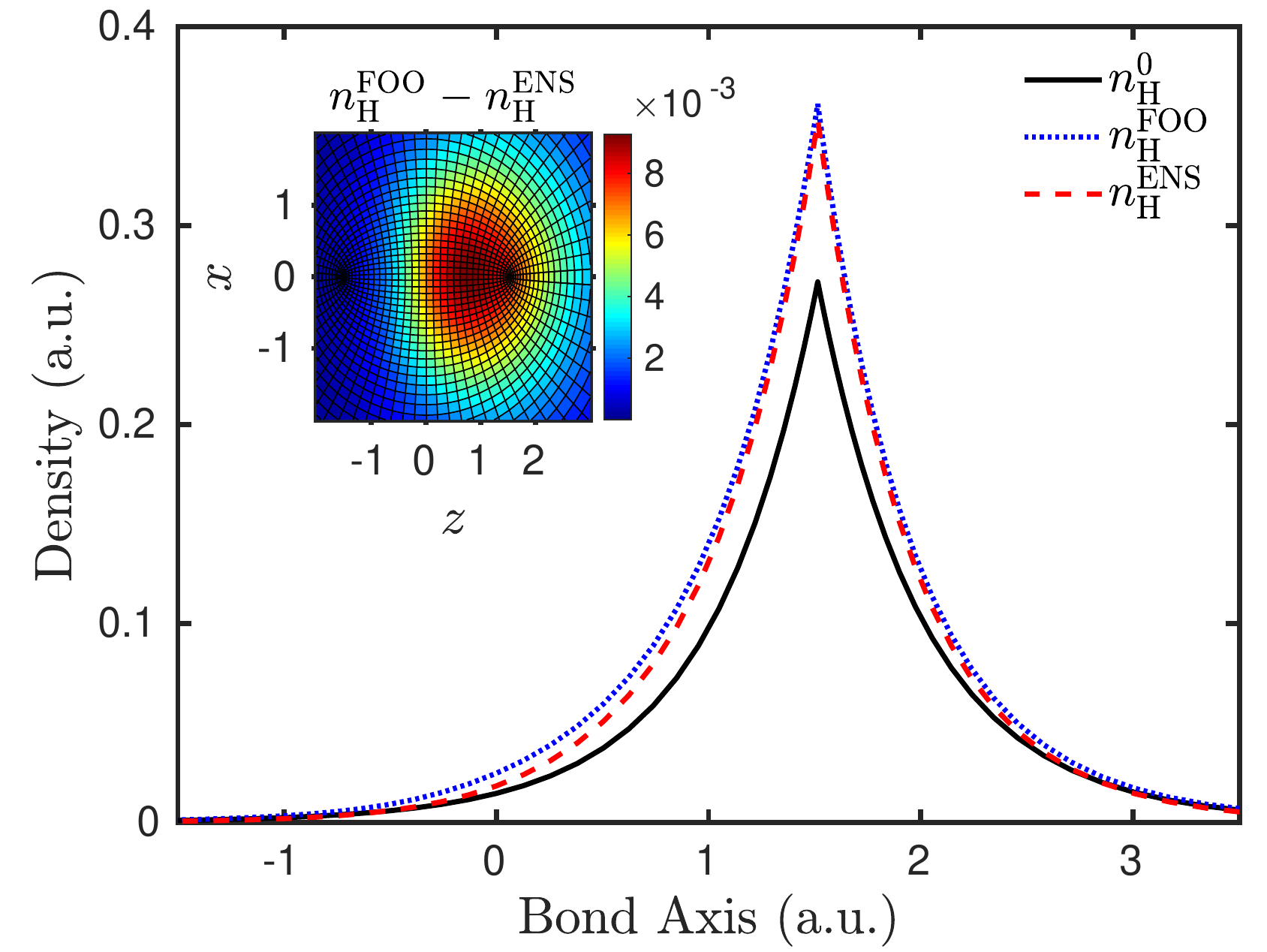}
	\caption{Optimal electron densities for the $\mathrm{H}$ fragment in $\mathrm{LiH}$ with FOO and ENS at the equilibrium internuclear distance $R=3.03$ a.u.; the solid line corresponds to an isolated hydrogen atom density.  The inset shows the difference between the FOO and ENS fragment densities on the $xz$-plane.}
	\label{fig:H_density-in-LiH}
\end{figure}

\begin{figure}[htp]
	\centering
	\includegraphics[width=1.0\linewidth]{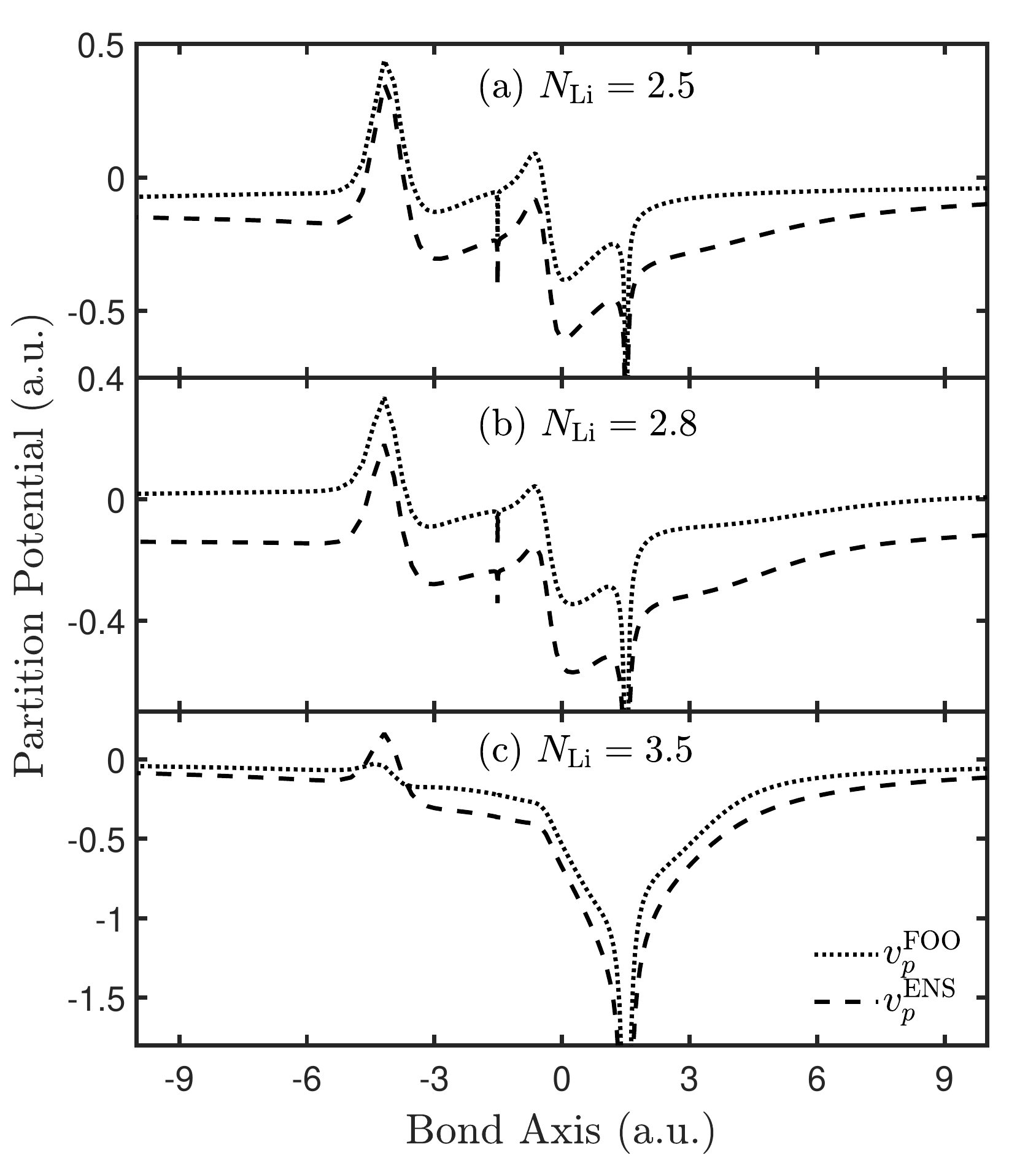}
	\caption{ Partition potential along the bond axis 
	for $\mathrm{LiH}$ with FOO and ENS at the equilibrium internuclear distance $R=3.03$ a.u. when $N_{\mathrm{Li}}=$ (a) 2.5, (b) 2.8, and (c) 3.5.}
	\label{fig:LiH_vp}
\end{figure}

Finally, we briefly examine the 4-electron molecule LiH at its equilibrium separation ($R_{eq}=3.03$ a.u.). In previous work
\cite{nafziger_molecular_2011}, it was speculated that 
the optimal P-DFT electron populations of the atomic fragments would be non-integers at $R_{eq}$.
However, a recent {\em exact} P-DFT calculation\cite{oueis_exact_2018} of a one-dimensional two-electron model of $\mathrm{LiH}$
yielded strictly integer populations.  
Figure \ref{fig:LiH} shows our LDA P-DFT energies for 3D $\mathrm{LiH}$ using both ENS and FOO. 
We find that, using ENS, the populations are indeed integers, as in the one-dimensional two-electron model \cite{oueis_exact_2018}.
Although the densities of $\mathrm{Li}$ and $\mathrm{H}$ fragments are distorted versions of the corresponding isolated-atom densities (see the H-atom density distortions in Fig.\ref{fig:H_density-in-LiH}), 
there is no electron transfer between fragments in ENS.
However, using FOO, 
the fragment electron populations are non-integral.
At the equilibrium internuclear distance, $R_{eq}=3.03$ a.u., 
we obtain $N_{\mathrm{Li}}^*=2.663$ and $N_{\mathrm{H}}^*=1.337$.
%Therefore, not only the densities of $\mathrm{Li}$ and $\mathrm{H}$ fragments are distorted, 
%but also electrons are transferred between fragments.
The number of electrons transferred from $\mathrm{Li}$ to $\mathrm{H}$ is $\Delta N=0.337$.
Just as in the examples of Secs.III-A and III-B, the number of electrons transferred with the FOO method is larger than with the ENS method (which is just zero in this case). 
To understand why, we distinguish three zones and note that when the number of electrons in the lithium atom is less than the optimum predicted by FOO ({\em zone 1}: $N_{\mathrm{Li}}<2.663$) then $\epsilon_{\mathrm{H},\mathrm{FOO}}^{\mathrm{HOMO}}=\epsilon_{\mathrm{H},\mathrm{ENS}}^{\mathrm{HOMO}}=\epsilon_{\mathrm{LiH}}^{\mathrm{HOMO}}$. However, 
when that number is higher than the optimum predicted by ENS ({\em zone 2}: $N_{\mathrm{Li}}>3.0$) then
$\epsilon_{\mathrm{Li},\mathrm{FOO}}^{\mathrm{HOMO}}=\epsilon_{\mathrm{Li},\mathrm{ENS}}^{\mathrm{HOMO}}=\epsilon_{\mathrm{LiH}}^{\mathrm{HOMO}}$. Finally, when $N_{\mathrm{Li}}$ is in between those two values ({\em zone 3}: $2.663<N_{\mathrm{Li}}<3.0$) then
$\epsilon_{\mathrm{Li},\mathrm{FOO}}^{\mathrm{HOMO}}=\epsilon_{\mathrm{H},\mathrm{ENS}}^{\mathrm{HOMO}}=\epsilon_{\mathrm{LiH}}^{\mathrm{HOMO}}$ and
$\epsilon_{\mathrm{Li},\mathrm{ENS}}^{\mathrm{HOMO}}<\epsilon_{\mathrm{LiH}}^{\mathrm{HOMO}}$.
Our analysis of Sec.II-C then implies that $v_{p}^{\mathrm{ENS}}\left(\mathbf{r}\right)$ is deeper than $v_{p}^{\mathrm{FOO}}\left(\mathbf{r}\right)$ (Fig.\ref{fig:LiH_vp}) and
$E_{f}^{\mathrm{ENS}}>E_{f}^{\mathrm{FOO}}$ (Fig.\ref{fig:LiH}). It can be seen in Fig.\ref{fig:LiH_vp} that the partition potential does not change qualitatively when $N_{\mathrm{Li}}$ crosses from zone 1 to zone 3, but there are major qualitative changes when it crosses to zone 2.
%%To understand why, we note that when the number of electrons in the lithium atom is less than the optimum predicted by FOO ($N_{\mathrm{Li}}<2.663$)  then $\epsilon_{\mathrm{Li},\mathrm{FOO}}^{\mathrm{HOMO}}=\epsilon_{\mathrm{Li},\mathrm{ENS}}^{\mathrm{HOMO}}=\epsilon_{\mathrm{LiH}}^{\mathrm{HOMO}}$,
%%and when that number is higher than the optimum predicted by ENS ($N_{\mathrm{Li}}>3.0$) then
%%$\epsilon_{\mathrm{H},\mathrm{FOO}}^{\mathrm{HOMO}}=\epsilon_{\mathrm{H},\mathrm{ENS}}^{\mathrm{HOMO}}=\epsilon_{\mathrm{LiH}}^{\mathrm{HOMO}}$.
%%Our analysis of Sec.II-C implies that, in those ranges of $\{N_{\mathrm{Li}}\}$, $v_{p}^{\mathrm{ENS}}\left(\mathbf{r}\right)$ is deeper than  $v_{p}^{\mathrm{FOO}}\left(\mathbf{r}\right)$ (Fig.\ref{fig:LiH_vp}) and
%%$E_{f}^{\mathrm{ENS}}>E_{f}^{\mathrm{FOO}}$ %%(Fig.\ref{fig:LiH}).
%%Interestingly, Figs.\ref{fig:LiH}-\ref{fig:LiH_vp} also show that the same conclusion holds even in the range $2.663<N_{\mathrm{Li}}<3.0$, where our analysis of Sec.II-C does not apply because $\epsilon_{\mathrm{H},\mathrm{FOO}}^{\mathrm{HOMO}}=\epsilon_{\mathrm{Li},,\mathrm{ENS}}^{\mathrm{HOMO}}=\epsilon_{\mathrm{LiH}}^{\mathrm{HOMO}}$. To what extent these observations hold in more general cases remains an open question.

%%%%%%%%
%Conclusion %
%%%%%%%%
% \newpage
\section{Concluding remarks}

% Kui: You want to add a paragraph stressing the distinction between dipole moments and charge transfer in the FOO vs. ENS comparison. Please send me a paragraph labeled "D". 

By treating fragment electron populations as variables in P-DFT calculations, we have shown how to find optimal populations via two alternative methods, one that involves fractional orbital occupations (FOO), and another that makes use of ensemble averages (ENS).  The optimal populations are found in both cases when the sum of fragment energies $E_f$ reaches its global minimum and all fragment 
electronegativities become equal. At that optimum, a value for the charge transferred between two fragments can be assigned unambiguously even at finite internuclear separations.  
%When $E_f[\{N_\alpha\}]$ does not display a cusp at integer numbers (locking the fragments into integer populations), our numerical results show that the condition of electronegativity equalization provides a sharp criterion for finding the minimum.  This criterion is in fact sharper than direct energy minimization, as the flatness of $E_f$ often poses a challenge to locate the minima.

%We also compare FOO and ENS in P-DFT by studying the fragment electronegativities, 
%optimal electron populations, and electric dipole moments.
%In FOO, there always exist one fragment whose electronegativity is a constant.
%However, fragment electronegativities change with electron populations in ENS.
%For integer electron populations in ENS, we find a criterion to determine if 
%it is a global minimizer of $E_f$.
%The number of transferred electrons and fragment dipole moments 
%given by FOO are larger than those given by ENS.

Through the formal analysis of Sec.II and the explicit numerical calculations of Sec.III, we have revealed differences between FOO and ENS that had not been observed in previous studies:
{\bf (1)} Although both methods lead to the same molecular densities and energies, the fragment densities of hetero-nuclear diatomic molecules can be significantly different, with the 
FOO method consistently yielding a larger fraction of an electron transferred from donor to acceptor, at least when the LDA is employed; {\bf (2)} The FOO fragment dipole moments are observed to provide an upper bound to the corresponding ENS dipole moments;
{\bf (3)} The ENS partition potentials are deeper than the corresponding FOO partition potentials; {\bf (4)} Accordingly, $E_{f}^{\mathrm{ENS}}>E_{f}^{\mathrm{FOO}}$ and $E_{p}^{\mathrm{ENS}}<E_{p}^{\mathrm{FOO}}$.  

Although a general proof of these observations is far beyond the scope of the present work, we suspect that FOO will generally transfer more electrons than ENS: When the nuclei are far enough apart, the effect of the partition potential on the fragment densities is negligible and the ENS fragment energies vary linearly with electron number. Since this occurs for both fragments, and since the FOO energies are observed to be convex functions of the electron number, the sum of fragment energies is expected to be lower in FOO than in ENS. The situation is less clear at finite internuclear separations $R$, but the small deviations from linearity detected for ENS even at small $R$ (certainly at equilibrium), suggest that the conclusions will remain valid in general at finite $R$.  
%*** TAlk about evolution of the cusp in ENS ***

%For approximate XC functionals, P-DFT with FOO and ENS partition the same molecular density differently.
%In order to show the differences between $n_{\alpha}^{\mathrm{FOO}}\left(\mathbf{r}\right)$
%and $n_{\alpha}^{\mathrm{ENS}}\left(\mathbf{r}\right)$,
%we can plot $n_{\alpha}^{\mathrm{FOO}}\left(\mathbf{r}\right)-n_{\alpha}^{\mathrm{ENS}}\left(\mathbf{r}\right)$ directly.
%And we also can define fragment electron populations $N_{\alpha}=\int n_{\alpha}\left(\mathbf{r}\right)d\mathbf{r}$ and transferred electrons
%to show the differences.
%For homo-nuclear diatomic molecules, FOO and ENS always give the same
%fragment electron population and transferred electrons.
%However, for hetero-nuclear diatomic molecules,
%FOO gives larger number of transferred electrons than that of ENS.
%For diatomic molecules, we also can use fragment electric dipole moments $\mathbf{p}_{\alpha}=\int\mathbf{r}n_{\alpha}\left(\mathbf{r}\right)d\mathbf{r}$
%since they are along the z direction.
%And we find that $\left|\mathbf{p}_{\alpha}^{\mathrm{FOO}}\right|>\left|\mathbf{p}_{\alpha}^{\mathrm{ENS}}\right|$.

%Differences between P-DFT with FOO and ENS:

%Adam: Change populations to populations throughout.

With appropriate approximations to the partition energy functional, P-DFT has been recently shown to overcome static-correlation and delocalization errors when stretching homonuclear diatomic molecules \cite{nafziger_fragment-based_2015}. Also for homonuclear diatomic molecules, a proposed ``covalent" approximation for the non-additive non-interacting kinetic energy shows promising results for orbital-free calculations \cite{jiang_constructing_2018}. An extension of these approximations to the ionic case and, more generally, to chemical bonds connecting inequivalent fragments, would be desirable. This extension will require choosing a method to treat fractional populations. Although both FOO and ENS methods are valid candidates, the results of this work point to advantages of each method in different cases: The observation that ENS fragment densities are less distorted than FOO densities upon formation of a chemical bond is a significant advantage for ENS, both conceptually and computationally. Conceptually, it is pleasing to have fragments-in-molecules that are minimally distorted from their isolated counterparts. Computationally, although each iteration of the P-DFT equations involves two integer-number calculations ({\em vs.} only one in FOO), the minimal distortion of the isolated input densities leads to a faster convergence that will benefit future applications. On the other hand, the FOO method has an advantage over ENS at equilibrium separations, as fractional populations are more in line with the chemist's intuition that polar molecules are composed of fragments with fractional formal charges. 

%We formulate partition density functional theory with constrained fragment electron populations where we take electron populations as parameters. 
%In this way, 

%For heteronuclear diatomic molecules, FOO gives non-integer optimal electron 
%populations, while ENS can give integer electron populations and the number 
%of transferred electrons are zero for certain molecules.
%We also can use P-DFT to describe the dissociation process of heteronulcear 
%diatomic molecules.

% (1) We find that ENS gives smaller transferred electrons than FOO.
% Furthermore, the electrons transferred in ENS is zero for certain molecules.

% The absolute values of dipole moments of fragment densities given by FOO 
% is bigger than those of ENS.

% (2)
% FOO always gives non-integral electron populations for heteronuclear diatomic molecules (can you prove it?), 
% while ENS gives integral electron populations for certain molecule.

% (3) For integer electron populations, ENS gives discontinuous electronegativites,
% when it is the minimizer of $E_f$?

% (4) if the total density is accurate, P-DFT can give the good picture of dissociation.

% H2: homonuclear, integral electron populations for finite and infinite distance

% HeH+: heteronuclear, integral electron populations for infinite distance

% LiH: heteronuclear, integral electron populations for finite and infinite distance

%%%%%%%%%%%%
%Acknowledgements %
%%%%%%%%%%%%
\begin{acknowledgments}
The authors thank Yan Oueis for valuable discussions.
This material is based upon work supported by the National Science Foundation 
under Grant No. CHE-1900301.

\end{acknowledgments}

%\section*{AUTHOR DECLARATIONS}
%\subsection*{Conflict of Interest}
%
%The authors have no conflicts to disclose.
%
%\section*{Data Availability}
%
%The data that support the findings of this study are available from the 
%corresponding author upon reasonable request.

% \nocite{*}
\bibliography{pdftLib}% Produces the bibliography via BibTeX.

\end{document}